\documentclass[journal]{IEEEtran}

\usepackage{cite}
\usepackage{amsmath,amssymb,amsfonts}
\usepackage{algorithmic}

\usepackage{graphicx}
\usepackage{textcomp}
\usepackage{xcolor}

\usepackage[ruled,lined,linesnumbered]{algorithm2e}
\usepackage{amsmath}
\usepackage{mathtools}

\usepackage{multirow}
\usepackage{multicol}

\usepackage{eurosym}
\usepackage{pifont}
\usepackage{caption}
\usepackage{subcaption}
\usepackage{adjustbox}

\usepackage{tikz,pgfplots}
\usepackage{pgfplotstable}
\pgfplotsset{compat=1.3}
\usetikzlibrary{patterns}
\usepackage{amsfonts}
\usepackage{cite}
\usepackage{url}
\usepackage{booktabs}
\usepackage{lineno}
\usepackage{amsmath}
\usepackage{multirow}
\usepackage[T1]{fontenc}
\usepackage{booktabs}
\usepackage{array}
\usepackage{chngpage}
\usepackage{tikz,pgfplots}
\usetikzlibrary{pgfplots.groupplots}
\usepackage{pgfplotstable}
\usetikzlibrary{patterns}
\usepackage{changes}
\definechangesauthor[name={Giacomo},color=blue]{gv}
\definechangesauthor[name={Giacomo},color=blue]{ql}
\definechangesauthor[name={Francesco},color=red]{fm}

\usepackage{soul}
\usepackage{todonotes}
\usepackage{graphicx}
\usepackage{subcaption}

\begin{document}
	
	\title{Multi-Failure Localization in High-Degree ROADM-based Optical Networks using Rules-Informed Neural Networks}
	
	\author{Ruikun Wang, \IEEEmembership{Graduate Student Member,~IEEE},
		Qiaolun Zhang,  \IEEEmembership{Graduate Student Member,~IEEE},
		Jiawei Zhang,   \IEEEmembership{Member,~IEEE},
		Zhiqun Gu,  \IEEEmembership{Member,~IEEE},
		Memedhe Ibrahimi, \IEEEmembership{Member,~IEEE},
		Hao Yu, \IEEEmembership{Member,~IEEE},
		Bojun Zhang,
		Francesco Musumeci, \IEEEmembership{Senior Member,~IEEE},
		Yuefeng Ji, \IEEEmembership{Senior Member,~IEEE},
		Massimo Tornatore, \IEEEmembership{Fellow,~IEEE}
		
		\thanks{This work was supported by National Key Research and Development Program of China (2022YFB2903700); National Natural Science Foundation of China (62271078); European Union under the Italian National Recovery and Resilience Plan (NRRP) of NextGenerationEU, Partnership on ``Telecommunications of the Future'' (PE00000001, program ``RESTART''). 
			(The first two authors contributed equally to this work.)  (Corresponding author: Jiawei Zhang; Yuefeng Ji.)
		}
		
		\thanks{Ruikun Wang, Jiawei Zhang, Zhiqun Gu, Bojun Zhang, and Yuefeng Ji are with the State Key Lab of Information Photonics and Optical Communications, Beijing University of Posts and Telecommunications, 100876 Beijing, China. 
			(email: wrk@bupt.edu.cn, zjw@bupt.edu.cn, guzhiqun@bupt.edu.cn, zbj@bupt.edu.cn, jyf@bupt.edu.cn)
		}
		
		\thanks{Qiaolun Zhang, Memedhe Ibrahimi, Francesco Musumeci, and Massimo Tornatore are with the Department of Electronics, Information and Bioengineering, Politecnico di Milano, 20133 Milano, Italy. (email: qiaolun.zhang@polimi.it, memedhe.ibrahimi@polimi.it, francesco.musumeci@polimi.it, massimo.tornatore@polimi.it)
		}
		
		\thanks{Hao Yu is with ICTFicial Oy, 02130, Espoo, Finland. (hao.yu@ictficial.com)
		}
		
	}
	
	\markboth{IEEE Journal on Selected Areas in Communications,~Vol.~X, No.~Y, \today}
	{Shell \MakeLowercase{\textit{et al.}}: A Sample Article Using IEEEtran.cls for IEEE Journals}
	
	
	\maketitle
	
	\begin{abstract}
		To accommodate ever-growing traffic, network operators are actively deploying high-degree reconfigurable optical add/drop multiplexers (ROADMs) to build large-capacity optical networks. 
		High-degree ROADM-based optical networks have multiple parallel fibers between ROADM nodes, requiring the adoption of ROADM nodes with a large number of inter-/intra-node components. 
		However, this large number of inter-/intra-node optical components in high-degree ROADM networks increases the likelihood of multiple failures simultaneously, and calls for novel methods for accurate localization of multiple failed components.  To the best of our knowledge, this is the first study investigating the problem of multi-failure localization for high-degree ROADM-based optical networks. To solve this problem, we first provide a description of the failures affecting both inter-/intra-node components, and we consider different deployments of optical power monitors (OPMs) to obtain information (i.e., optical power) to be used for automated multi-failure localization. Then, as our main and original contribution, we propose a novel method based on a rules-informed neural network (RINN) for multi-failure localization, which incorporates the benefits of both rules-based reasoning and artificial neural networks (ANN). Through extensive simulations and experimental demonstrations, we show that our proposed RINN algorithm can achieve up to around 20\% higher localization accuracy compared to baseline algorithms, incurring only around 4.14 ms of average inference time. 
	\end{abstract}
	
	
	\section{Introduction}
	\label{section-introduction}
	
\begin{figure}[htb]
		\centering    \includegraphics[width=0.95\linewidth]{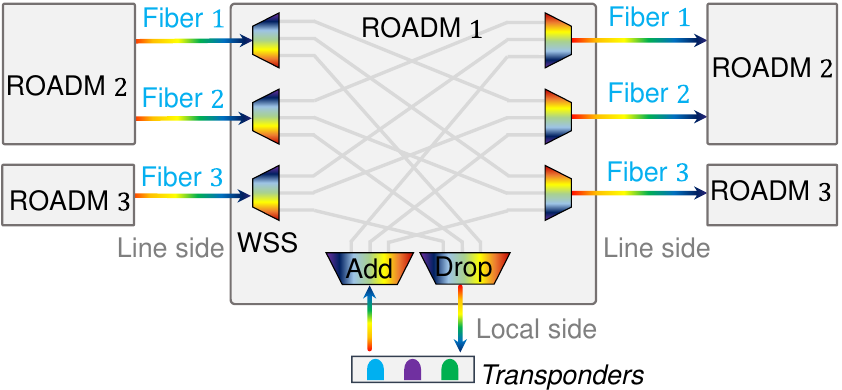}
		\caption{An example of a ROADM with degree 4.}
		\label{Fig-Concept_Degree_ROADM}
	\end{figure}

	\IEEEPARstart{O}ptical networks constitute the underlying ultra-large capacity platform to accommodate the massive traffic demands~\cite{9606720,brunner20236th} that will be required by emerging applications in Beyond-5G/6G networks, including metaverse~\cite{10413955}, digital twin~\cite{10198573}, and distributed edge computing~\cite{10002713,10.1145/3672202.3673744}. 
	Network operators are actively exploring how to build large-capacity optical networks that can simultaneously support multiple fiber links between network nodes and several nodal directions, by leveraging high-degree reconfigurable optical add/drop multiplexers (ROADMs)~\cite{li2022colorless,mehrvar2023dimensioning}. 
	The degree of a ROADM node (defined as \textit{ROADM degree}) is the sum of the number of outgoing fibers connecting the ROADM node to other neighboring ROADM nodes, plus the number of fibers connecting the ROADM node to local wavelength selective switches (WSSs) in the drop plane of the local side\footnote{For sake of clarity, in this work, we consider bidirectional networks with two fibers per degree. Under this assumption, the degrees in terms of entering fibers ($degree_{IN}$) and exiting fibers ($degree_{OUT}$) are the same, and in the rest of the paper we will consider, without loss of generality, always $degree_{OUT}$.}.
	For example, Fig.~\ref{Fig-Concept_Degree_ROADM} illustrates the case of a ROADM with degree 4. In fact, on the line side, $ROADM_1$ is directly connected to $ROADM_2$ through two outgoing fiber links, and to $ROADM_3$ through one outgoing fiber link; on the local side, two local WSSs are used as add plane and drop plane, respectively.  Thus, the ROADM degree of $ROADM_1$ is 4, as the sum of the number of outgoing fibers from $ROADM_1$ to all adjacent nodes (i.e., $ROADM_2$ and $ROADM_3$) and the number of drop plane of $ROADM_1$. 
	The need for a higher network throughput is leading network operators to investigate alternatives solutions such as multi-fiber or multi-band transmission~\cite{Sticca-23}. Multi-fiber networks with high-degree ROADMs~\cite{Oleg24MultiFiber} are currently considered as the most likely short- to medium-term solution, especially as high-degree ROADM nodes (supporting up to 32 outgoing fibers in a single ROADM node) are now becoming commercially available~\cite{huawei-P32-WSS}. 
	
	High-degree ROADMs typically comprise multiple components, such as WSSs, pre-amplifiers (Preamp), and booster amplifiers (Booster). These components result in intricate internal compositions within each ROADM node, forming \textit{intra-node networks}.  Meanwhile, the connection between two high-degree ROADMs involves multiple fiber spans and inline amplifiers (ILA), referred to as \textit{inter-node networks}. In high-degree ROADM-based optical networks, failures may occur in various components in both inter-/intra-node networks, leading to data loss and performance degradation. Moreover, since the likelihood of multiple components failing simultaneously increases as the degree of ROADM nodes grows, multi-failure localization for inter-/intra-node components becomes increasingly important to ensure reliable network management. 
	For example, the ability to localize intra-node failures enables new forms of intra-node lightpath recovery~\cite{flex-scale-eu} based on the surviving \lq \lq partial degree'' within a ROADM node, and it allows to substitute/repair only selected sub-components of a ROADM node rather than the whole ROADM node.
	
	On the other hand, multi-failure localization covering both inter-/intra-node components in high-degree ROADMs-based networks presents significant new challenges: ($i$) \textit{the number of failures is not a fixed value}, hence, compared to single-failure localization, multi-failure localization must address the ambiguity regarding the identification of number of simultaneous faulty components; ($ii$) \textit{the inter-/intra-node networks contain an extensive number of components}, which significantly complicates the task of pinpointing failure locations with precision.
	For instance, if a ROADM node has four adjacent ROADM nodes with 8 incoming/outgoing fibers to each adjacent ROADM node, then the overall number of inter-/intra-node components, namely incoming/outgoing fibers, pre-/booster/inline-amplifiers and WSS, will increase of factor 4 and 8, from 8 to 256\footnote{The factor 4 indicates the number of adjacent ROADM nodes, and the factor 8 indicates the number of degrees from the current ROADM node to an adjacent ROADM node. The value 8 denotes the number of components in each degree, which includes one incoming fiber, one outgoing fiber, one pre-amplifier, one booster-amplifier, two inline-amplifiers, and two WSS. Therefore, the total number of components is $4 \times 8 \times 8 = 256$.}.
	
	
	The main contributions of this paper can be summarized as follows:
	\begin{itemize}
		\item To the best of our knowledge, this is the first study investigating multi-failure localization in high-degree ROADM-based optical networks, considering both inter-/intra-node components. 
		\item We provide a description of the failures in high-degree ROADM-based optical networks, and we consider a set of different strategies for the deployment of optical power monitors (OPMs) for monitoring optical power. 
		\item As our main original contribution, we develop a new method based on rules-informed neural network (RINN) for multi-failure localization of inter-/intra-node components, which combines the advantages of both rules-based reasoning and artificial neural networks (ANN). 
		\item We extensively evaluate the performance of our proposed RINN algorithm with both simulations and experimental evaluations, demonstrating that our proposed RINN algorithm can achieve up to around 20\% higher localization accuracy compared to the baseline algorithms, incurring only around 4.14 ms of average inference time. 
	\end{itemize}
	
	The remainder of the paper is organized as follows.
	Section \ref{section-related-work} reviews the related work of this paper.
	Section \ref{section-problem-description} describes the multi-failure localization problem.
	Section \ref{section-rules-informed-neural-networks} introduces the proposed framework and dataset generator in detail. 
	In Section \ref{section-experimental-setup-and-results}, we provide simulation evaluations and experimental demonstrations of the proposed approach.
	Finally, Section \ref{section-conlusion} concludes the paper and discusses  future research directions.

	\section{Related Work}
	\label{section-related-work}
	
	This section reviews the related work on both single-failure and multi-failure localization in (ROADM-based) optical networks, and elaborates on the limitations of existing methods. Note that failures in optical networks can be classified as hard failures or soft failures \cite{mayer2021machine}, and our proposed methodology is applicable for localizing both of them.
	
	\subsection{Single-Failure Localization in Optical Networks}
	Failure localization has attracted significant research effort because it is vital to ensure swift network recovery~\cite{ishigaki2020deeppr,zhang2021progressive}. Many studies have addressed hard-/soft-failure localization of single failures occurring in fiber links or switching nodes in optical networks~\cite{wen2005efficient,zeng2006novel,wu2010optical,yang2020accurate,mayer2021machine,li2021fault,silva2022learning,Zeng_2023_10058051}. 
	For example, failure localization is carried on using heuristic algorithms and/or integer linear programming (ILP) models considering mainly one link failure~\cite{wen2005efficient,zeng2006novel,wu2010optical}.  
	Specifically, normal links can be easily identified if a receiver receives a normal optical signal, as all the links traversed by the corresponding lightpath (LP) are normal. Conversely, an abnormal link may be identified if multiple receivers of LPs crossing that link observe an abnormal optical signal~\cite{wu2010optical}. 
	
	With the recent rise of machine learning (ML) application, ML algorithms are being increasingly adopted for failure localization in optical networks. 
	Ref.~\cite{yang2020accurate} leverages neural evolution networks for fault localization in the optical interconnect network for cloud data centers. 
	Since failures might also affect telemetry, making available only a subset of the required telemetry data, Ref.~\cite{mayer2021machine} proposes an ML framework for localizing soft failures with partial telemetry. 
	Ref.~\cite{li2021fault} leverages knowledge graphs to analyze alarms, and introduces graph neural networks to locate the network faults.
	All the previously mentioned works focus on failure localization on either nodes and links, while Ref.~\cite{Zeng_2023_10058051, Wang_2023_10138320, wang2023meta,10477618} are the first to investigate failure localization for both intra-/inter-node components using ML-based approaches. 
	For instance, Ref.~\cite{Zeng_2023_10058051} proposed an attention-mechanism-enabled fault location method for both intra-/inter-node components based on network-wide monitoring data in broadcast-and-selected (B\&S) ROADM optical networks.
	However, these previous works did not consider multi-failure localization and could not be scaled to high-degree ROADM-based optical networks, while our RINN allows to tackle scenarios with large WSSs and the number of fibers per link up to 15. 
	
	Note that a novel monitoring technique, namely, power profile estimation (PPE), is under initial experimental demonstration~\cite{sasai2023performance,may2023longitudinal}, which has (not yet investigated) potential to perform failure localization on a network scale as discussed in Ref.~\cite{zhang2023cost}. 
	However, this work does not consider PPE for failure localization as the investigation of PPE is still at an early stage, and it has not yet been deployed in commercial optical networks.
	
	\subsection{Multi-Failure Localization in Optical Networks}
	While most literature focuses on single-failure localization, a few works have addressed multi-failure localization in optical networks using both heuristic algorithms~\cite{Luo_2012_6476364,xuan2013efficient,9203029,delezoide2023streamlined}. 
	Ref.~\cite{Luo_2012_6476364} presents a novel multi-fault localization mechanism utilizing a fuzzy fault set for efficient and scalable fault localization in large-capacity optical transport networks. 
	Ref.~\cite{xuan2013efficient} proposes two novel algorithms, namely, a tree-decomposition-based algorithm and a random walk-based algorithm for multi-link failure localization in all-optical networks.
	Ref. \cite{9203029} localizes soft failures by leveraging degradation detection of quality of transmission (QoT).
	\textcolor{black}{Ref.~\cite{delezoide2023streamlined} is limited to hard failures, while the proposed RINN solution extends to soft failures.
	}

	In addition to the aforementioned approaches based on heuristic algorithms, some other works also investigate ML-based algorithms for multi-failure localization~\cite{yang2019efficient,xin2021dnn,9763325,lun2021roadm}. Specifically, Ref.~\cite{yang2019efficient} proposes an efficient hybrid multi-failure localization algorithm based on Hopfield Neural Networks (HNNs) for coexisting radio and optical wireless networks in 5G. Ref.~\cite{xin2021dnn} utilizes both the backpropagation (BP) neural network and the long short-term memory (LSTM) neural network to enhance multi-fault localization in 5G coexisting radio and optical wireless networks. 
	Ref. \cite{9763325} reports the experimental demonstration of ML-based double-failure localization, where principal component analysis (PCA) is used to reduce non-essential information. 
	Ref.\cite{lun2021roadm} uses artificial neural network (ANN) and Gaussian process regression (GPR) to localize the soft-failure location. The authors consider two soft-failure types of WSS, and the localization accuracy reaches 95\%.
	Furthermore, Ref.\cite{1563847} considers different hard failures, including fiber links and network components. The results allow to determine the availability of a wavelength connection subject to the transmission distance and the number of traversed hops. 
	
	In summary, while some studies have explored multi-failure localization in optical networks, to the best of our knowledge, no existing work has specifically addressed multi-failure localization considering both intra- and inter-node component failures and considering a scalable solving method suitable for high-degree ROADM-based optical networks.
	
	\section{System Model}
	\label{section-problem-description}
	In this section, we describe the architecture of high-degree ROADM-based optical networks, and provide a possible comprehensive description for different failure types of inter-/intra-node components.  We also introduce the OPM and deployment strategies.
	
	\subsection{High-Degree ROADM-based Optical Networks}
	
	\begin{figure}[tb]
		\centering
		\includegraphics[width=0.95\linewidth]{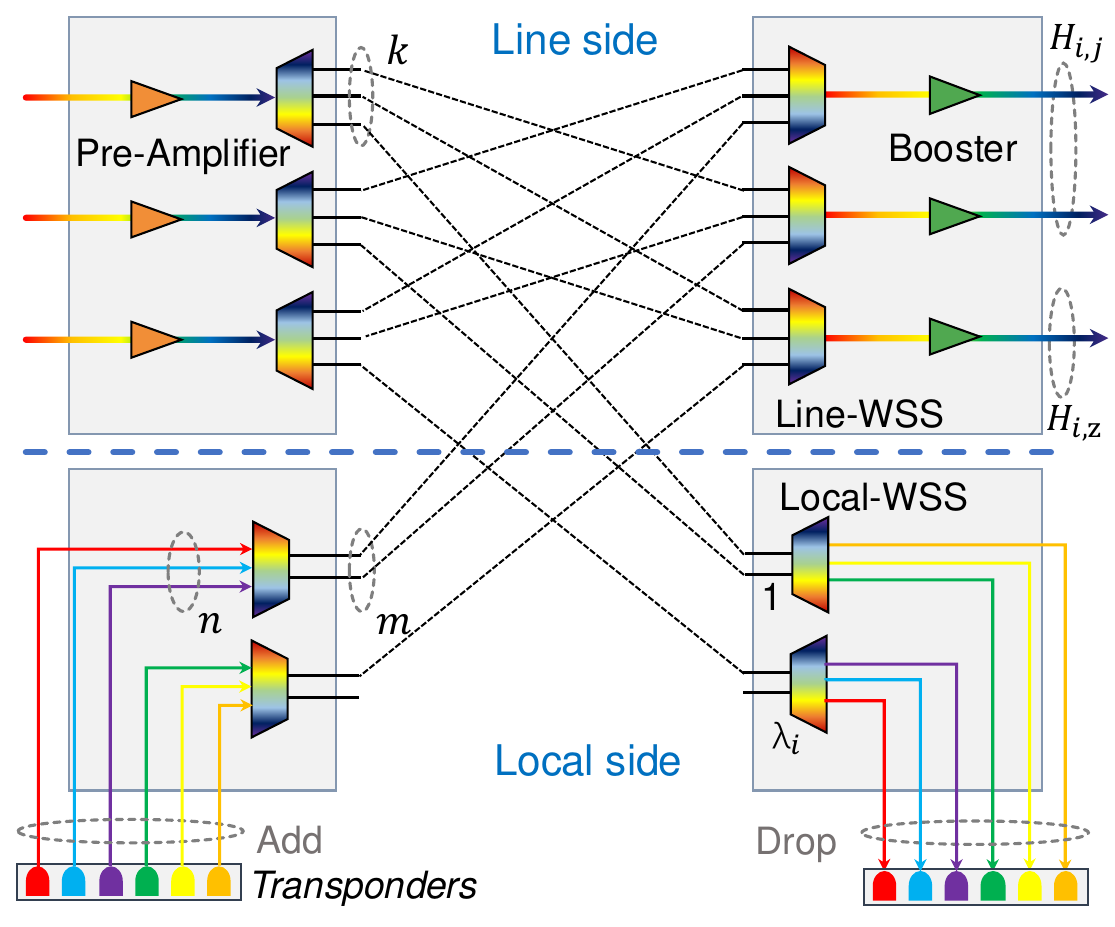}
		\caption{A fully-connected high-degree ROADM node.}
		\label{Fig-High-Degree-ROADM}
	\end{figure}
	
	High-degree ROADM-based optical networks consist of high-degree ROADM nodes interconnected by multiple fibers between adjacent ROADM node pairs. This work considers a fully-connected high-degree ROADM architecture, as shown in Fig.~\ref{Fig-High-Degree-ROADM}. 
	The ROADM node can support different directions connecting to other ROADM nodes (the \lq \lq line side'') together with one direction to the add/drop module in the node itself (the \lq \lq local side''). 
	Let us denote the set of neighboring ROADM nodes of ROADM node $i$ as $\mathcal{R}_i$. 
	Assume that the ROADM node $i$ can connect to the neighboring ROADM node $j \in \mathcal{R}_i$ with $H_{i,j}$ outgoing links, indicating that neighboring ROADM node $j$ contribute to $H_{i,j}$ degrees for ROADM node $i$. 
	Thus, the degree of ROADM node $i$ related to the line side is $\sum_{j \in \mathcal{R}_{i}} H_{i,j}$. 
	In addition, assume that the number of WSS in the drop plane is $\lambda_i$, then the degree of ROADM node $i$ related to the local side is $\lambda_i$. 
	Hence, the ROADM degree of node $i$ can be calculated with Eqn.~(\ref{eq:number-of-degree}), where $\mathcal{N}$ is a set of ROADM nodes. 
	Fig.~\ref{Fig-High-Degree-ROADM} shows an example of calculating the degrees of the ROADM node $i$, which has two neighboring ROADM nodes (node $j$ and node $z$). 
	The number of outgoing links from ROADM node $i$ to ROADM nodes $j$ and $z$ is 2 and 1, respectively, and the number of WSS on the local side is 2. 
	Hence, the ROADM degree of node $i$ is 5. 
	All the sets and parameters are listed in Table.~\ref{tab:parameters}.
	\begin{equation}
		L_{i} = \sum_{j \in \mathcal{R}_{i}} H_{i,j} + \lambda_{i}, \forall i \in \mathcal{N}
		\label{eq:number-of-degree}
	\end{equation}
	
	\begin{table}[htbp] 
		\small
		\caption{Sets and Defined Parameters}
		\centering
		\label{tab:parameters}
		\begin{tabular}{ll} 
			\toprule 
			Params & Description\\
			\midrule
			$\mathcal{N}$ & Set of ROADM nodes \\
			$\mathcal{L}$ & Set of LP requests\\
			$\mathcal{M}$ & Set of candidate OPM locations \\
			$\mathcal{R}_i$ & Set of neighboring ROADM nodes \\
			& of ROADM node $i$, $i \in \mathcal{N}$\\
			$\Psi_{m}$ & Binary, 1 if OPM is placed in \\
			& candidate location $m$, $m \in \mathcal{M}$\\
			$L_i$ & ROADM degree of node $i$, $i\in \mathcal{N}$\\
			$\zeta_{i,j}$ & Length of link $(i,j)$, $i,j \in \mathcal{N}, i \neq j$ \\
			$z$ & Components in all LPs \\
			$x$ & Power values of OPM before failure \\
			$\overline{x}$ & Power values of OPM after failure \\
			$C$ & Number of components\\ 
			$C_{node}$, $C_{link}$ & Number of node and link components \\
			$M$ & Number of candidate OPM locations, i.e., $|\mathcal{M}|$\\
			$M_{node}$, $M_{link}$ & Number of candidate OPM locations \\
			& in nodes and links\\
			$H_{i,j}$ & Number of single-direction fibers between \\
			& node $i$ and node $j$, $i, j \in \mathcal{N}, i \neq j$ \\
			$m$, $n$ & Number of ports of local-WSS ($m \leq n$) \\
			$k$ & Number of ports of the side for intra-node \\
			& connection in the considered $1 \times k$ line-WSS \\
			$S_{i,j}$ & Number of fiber spans in link $(i,j)$, \\
			& $i,j \in \mathcal{N}, i \neq j$\\
			$L^{span}$ & Length of single fiber span \\
			$\lambda_{i}$ & Number of local WSSs in add side \\
			& or drop side of ROADM node $i$, $i \in \mathcal{N}$\\
			$\alpha$ & A negative constant value for padding \\
			& OPM location when no OPM is \\
			& deployed in the location \\
			\bottomrule 
		\end{tabular} 
	\end{table}
	
	Let us now discuss the constraints of the number of ports of the WSSs in the ROADM node. The line side is equipped with $1\times k$ line-WSS, and the local side is equipped with $n\times m$ local-WSS ($n \geq m$). Since we consider a fully-connected ROADM architecture, the signal received in each line-WSS needs to have enough ports to directly connect to all the other line-WSSs and one local-WSS at the local side. 
	Hence, the number of ports (i.e., $k$) of the line-WSS for intra-node connection should satisfy the following Constraint~(\ref{eq-constraintLinePorts}):
	\begin{equation}
		k \geq \sum_{l \in \mathcal{R}_{i}, l \neq j} H_{i,l} + 1, \forall i \in \mathcal{N}, j \in \mathcal{R}_i
		\label{eq-constraintLinePorts}
	\end{equation}

	The ROADM degree is related to the connection between the line-WSSs on the line side and local-WSSs on the local side. 
	In this work, we consider that a line-WSS is only connected to a local-WSS. The remaining ports (equal to $k-1$) are connected to up to $k-1$ line-WSSs, and hence we can have $k$ line-WSSs. 
	Since the number of ports $m$ in $n \times m$ WSS is usually much smaller than the number of ports $k$ in $1 \times k$ WSS with the current devices~\cite{huawei-P32-WSS,8386355}, we need multiple local-WSSs to connect to all the line-WSSs. Specifically, the Constraint~(\ref{eq-constraintLocalPorts}) enforces that the local side can connect to any line-WSS. The total number of ports in all the local-WSSs for connection with the line side should be greater than the number of line-WSSs on the line side. According to Constraint~(\ref{eq-constraintLocalPorts}), we can deduct the number of local-WSS required on the local side as shown in Constraint~(\ref{eq-Nadi}), where $\lceil x \rceil$ denotes the ceiling operation for $x$.
	\begin{equation}
		m \cdot \lambda_{i} \geq \sum_{j \in \mathcal{R}_{i}} H_{i,j}, \forall i \in \mathcal{N}
		\label{eq-constraintLocalPorts}
	\end{equation}
	\vspace{-3pt}
	\begin{equation}
		\lambda_{i} = \left\lceil \dfrac{\sum_{j \in \mathcal{R}_i}{H_{i,j}}}{m} \right\rceil
		\label{eq-Nadi}
	\end{equation}
	
\begin{figure}[tb]
		\centering
		\includegraphics[width=1.0\linewidth]{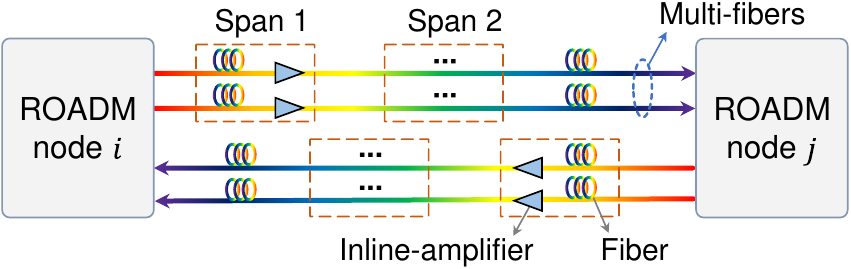}
		\caption{A multi-fiber link between two ROADM nodes.}
		\label{Fig-multi-fiber-link}
	\end{figure}

	Let us now enumerate all the possible failed components, which can be used to quantify the complexity of the failure localization problem for high-degree ROADM-based optical networks. The number of components $C$ in high-degree ROADM-based optical networks can be calculated as the sum of the number of intra-node components (denoted with $C_{node}$) and the number of link components  (denoted with $C_{link}$) as in Constraint~(\ref{eq-C}).  We ignore failures of \textit{intra-node fibers} and \textit{connectors} due to their typically minimal impact on the overall attenuation.
	The total number of intra-node components can be calculated as in Eqn.~(\ref{eq-Cnode}), which includes the number of transponders, local WSSs\footnote{A WSS with $m \times n$ ports is usually a sub-WSS system~\cite{8386355}, but for simplicity, this paper considers it as a single component.}, 
	pre-amplifiers/boosters, and line WSSs. 
	As shown in Fig. \ref{Fig-multi-fiber-link}, we consider multiple fibers to connect ROADM node $i$ and ROADM node $j$.
	As for the inter-node components, let us $S_{i,j}$ denote the number of fiber spans in link $(i,j)$, which can be calculated by Eqn.~(\ref{eq-num-of-span}) as the ceiling of the ratio of the length $\zeta_{i,j}$ of link $(i,j)$ over the length of a single span denoted with $L^{span}$. 
	Then the number of components in a single fiber from node $i$ to node $j$ is $2 \cdot S_{i,j} -1$, which includes $S_{i,j}$ fiber spans and $S_{i,j}-1$ in-line amplifiers. Thus, the total number of link components $C_{link}$ can be calculated as in Constraint~(\ref{eq-Clink}). 
	\begin{equation}
		C = C_{node} + C_{link}
		\label{eq-C}
	\end{equation}
	\vspace{-3pt}
	\begin{equation}
		C_{node} = \sum_{i\in \mathcal{N}} \left(
		n \cdot \lambda_{i} + 
		2\cdot \lambda_{i} + 
		4 \cdot \sum_{j \in \mathcal{R}_i}{H_{i,j}}
		\right)
		\label{eq-Cnode}
	\end{equation}
	\vspace{-3pt}
	\begin{equation}
		S_{i,j} =  \left\lceil \frac{\zeta_{i,j}}{L^{span}}  \right\rceil
		\label{eq-num-of-span}
	\end{equation}
	\vspace{-3pt}
	\begin{equation}
		C_{link} = \sum_{i \in \mathcal{N}} \sum_{j \in \mathcal{R}_i} H_{i,j} \cdot \left(2 \cdot S_{i,j} - 1 \right)
		\label{eq-Clink}
	\end{equation}
	\vspace{-3pt}
	
	\begin{table*}[htb]
		\caption{Description of Failure Types}
		\centering
		\begin{tabular}{l|l|l}
			\hline
			\textbf{Failure components}             & \textbf{Failure Types}   & \textbf{Failure Descriptions}                    \\ \hline
			\multirow{2}{*}{Transponders}           & Break (hard failure)                    & Launch power is lower than the break threshold   \\ \cline{2-3} 
			& Launch power degradation (soft failure) & Launch power is lower than the expected value     \\ \hline
			\multirow{2}{*}{Pre-/Post-/Line-amplifiers} & Break (hard failure) & Amplifier gain is lower than the break threshold \\ \cline{2-3} 
			& Gain degradation (soft failure)         & Amplifier gain is lower than expected value       \\ \hline
			\multirow{3}{*}{Local-WSS, line-WSS}    & Break (hard failure)                    & Insertion loss exceeds the break threshold       \\ \cline{2-3} 
			& Excessive filtering (hard failure)     & Abnormal wavelength filtering                    \\ \cline{2-3} 
			& Extra attenuation (soft failure)       & Insertion loss is larger than the expected value \\ \hline
			\multirow{2}{*}{Inter-node fiber spans} & Break (hard failure)                   & Fiber loss exceeds the break threshold           \\ \cline{2-3} 
			& Loss degradation (soft failure)         & Fiber loss is larger than expected loss          \\ \hline
		\end{tabular}
		\label{tab:failure_models}
	\end{table*}
	
	\subsection{Description of Failure Types}
	
	High-degree ROADM-based optical networks suffer from failures in four categories, including transponders, pre-/line-amplifiers, boosters, local-/line-WSS, and inter-node fiber spans as shown in Table \ref{tab:failure_models}. 
	We consider both hard failures, which completely disrupt the signal, and soft failures, which degrade the signal. 
	A common hard failure across multiple components is \textit{break}. This refers to a situation where the component is entirely not functional, usually because its operational parameters fall below a critical threshold required for functionality (defined as \textit{break threshold}). For transponders, a \textit{break} occurs when the signal launch power is too low to maintain transmission. For amplifiers, \textit{break} indicates that the amplifier gain is insufficient (lower than the break threshold) to amplify the signal properly. For inter-node fiber spans, a \textit{break} occurs when fiber loss is so high that it exceeds the break threshold, leading to a complete signal loss. Another hard failure considered is the \textit{excessive filtering} of local-WSS and line-WSS components in case of abnormal wavelength filtering. 
	Aside from hard failure, we also consider soft failures as follows. Transponders, pre-/line-amplifiers, and boosters may suffer from launch power degradation and gain degradation, respectively, which imply that the components are functioning, but at a reduced functionality. Specifically, \textit{launch power degradation} occurs in the transponder when the launch power is lower than the expected value, and \textit{gain degradation} occurs in the pre-/line-amplifiers and boosters when the amplifier gain is lower than the expected value. 
	Besides, local-WSS and line-WSS components can also experience \textit{extra attenuation} when the insertion loss exceeds the expected value. Moreover, the inter-node fiber spans suffer from loss degradation when the fiber loss is larger than the expected loss. In this study, we assume that all the failures listed in Table \ref{tab:failure_models} may occur simultaneously, and the objective is to localize and identify all the failures.

	\subsection{Optical Performance Monitors}
	
	\subsubsection{Monitoring of Signal Power}
	
	In this work, we consider deploying low-cost OPMs that only monitor signal power to reduce the monitoring cost.  Although expensive and more complex OPMs~\cite{saif2020machine} can monitor other physical metrics beyond signal power, e.g., BER, OSNR, etc, as they are usually expensive and are not considered in this work.

\begin{figure}[tb]
		\centering
		\includegraphics[width=1.0\linewidth]{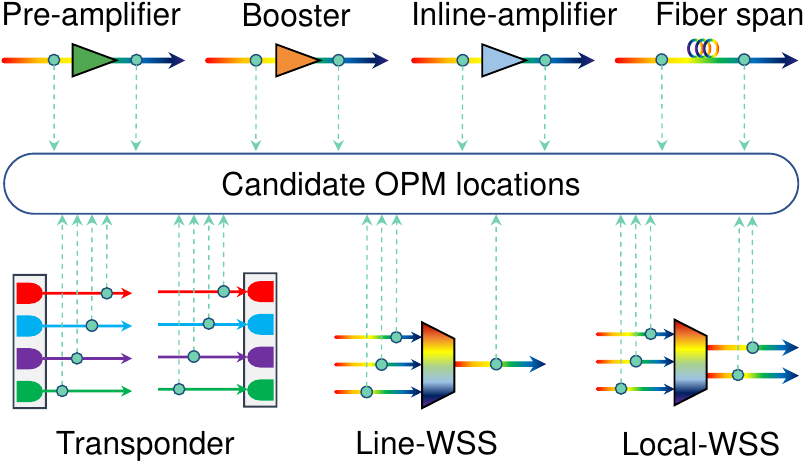}
		\caption{Candidate OPM locations.}
		\label{Fig-OPM-positions}
	\end{figure}

	OPMs can be deployed between every two adjacent components listed in Table~\ref{tab:failure_models}, and the location where OPM can be deployed is defined as \textit{candidate OPM location}. Assume that the total number of candidate OPM locations is denoted with $M$, which equals the sum of the number of candidate OPM locations in nodes (defined as $M_{node}$) and the number of candidate OPM locations in links (defined as $M_{link}$), as shown in Constraint~(\ref{eq-M}). Regarding the candidate locations of nodes, as shown in Constraint~(\ref{eq-Mnode}), OPMs can be placed in the input/output of pre-amplifiers/boosters, output (input) of the transponders, and all ports of line-WSS together with local-WSS as shown in Fig.~\ref{Fig-OPM-positions}. Specifically, the number of candidate OPM locations between transponders and local WSSs equals the number of transponders, namely $2 \cdot n \cdot \lambda_{i}$. The number of candidate OPM locations related to pre-amplifiers/boosters, either between pre-amplifiers/boosters and line-WSSs or between pre-amplifiers/boosters and inter-node fibers, is directly equal to the number of pre-amplifiers/boosters, denoted with $\sum_{j \in \mathcal{D}_i} (2 \cdot H_{i,j})$. Besides, the number of candidate OPM locations between local-WSSs and line-WSSs equals $\sum_{j \in \mathcal{D}_i} (2 \cdot H_{i,j})$, considering the add and drop plane. 
	In addition, the number of candidate OPM locations between line-WSSs equal $\sum_{j \in \mathcal{D}_i} \sum_{l \in \mathcal{D}_{i}, l \neq j} H_{i,j} \cdot H_{i,l}$, as each port should directly connect to all the other ports. In summary, the number of candidate OPM locations for links can be calculated in Constraint~(\ref{eq-Mnode}). Regarding the candidate OPM locations in links, as shown in Fig.~\ref{Fig-OPM-positions}, OPM can be placed at the input and output of line-amplifiers. 
	These locations correspond respectively to the ending of the previous fiber span and the beginning of the current fiber span, and hence, the fiber span can also be monitored with the OPM placed at the input and output of the line amplifiers. 
	Constraint~(\ref{eq-Mlink}) shows the number of OPM locations in links. 
	
	\begin{equation}
		M = M_{node} + M_{link}
		\label{eq-M}
	\end{equation}
	\vspace{-3pt}
	\begin{equation}
		\begin{split}
			M_{node} = \sum_{i\in \mathcal{N}} \sum_{j \in \mathcal{D}_i} (6 \cdot H_{i,j} &+ \sum_{l \in \mathcal{D}_{i}, l \neq j} H_{i,j} \cdot H_{i,l}) \\
			&+ \sum_{i\in \mathcal{N}} 2 \cdot n \cdot \lambda_{i}
		\end{split}
		\label{eq-Mnode}
	\end{equation}
	\vspace{-3pt}
	\begin{equation}
		M_{link} = \sum_{i \in \mathcal{N}} \sum_{j \in \mathcal{D}_i} 2 \cdot (S_{i,j} - 1)
		\label{eq-Mlink}
	\end{equation}
	
	\subsubsection{OPM deployment}
	\label{OPM-deployment-strategies}
	
	Network operators prefer expect to deploy fewer OPMs to reduce capital expenditures, i.e., OPMs are not deployed ``by default'' everywhere~\cite{mayer2021machine}.
	In addition, some devices, such as transponders, may be already equipped with built-in OPMs to monitor the output/input power. 
	Consequently, the failure-cause localization accuracy heavily depends on strategic OPM deployment. Assume that the set of all OPM placement locations is denoted with $\mathcal{M}=\{1,2,...,M\}$. This work considers a uniform distribution to deploy OPMs (optimized deployment is left as future work), and the number of deployed OPMs is $M'$. Then, one OPM is deployed every $I =  \left\lfloor \dfrac{M}{M'} \right\rfloor$ OPM candidate locations, and the set of locations deployed with OPMs can be defined as $\mathcal{M'}=\{mI |m \in 1..M' \}$. 
	Let $\Psi_{m}$ represent the placement of OPM in a candidate location $m$, such that $\Psi_{m} = 1$ indicates OPM is placed in location $m$, as  detailed in Constraint~(\ref{eq:phi}):
	\begin{equation}
		\Psi_{m} = 
		\begin{cases}
			1, & \text{if } m \in \mathcal{M'} \\
			0, & \text{else }
		\end{cases},
		\forall ~ m \in \mathcal{M}
		\label{eq:phi}
	\end{equation}
	
\begin{figure}[htb]
		\centering
		\includegraphics[width=1.0\linewidth]{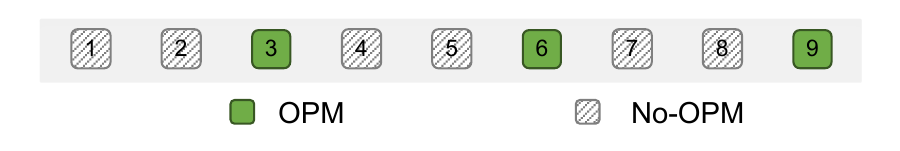}
		\caption{An example of uniform OPM deployment.}
		\label{fig:eg-deployment}
	\end{figure}

	We now show an example of uniform OPM deployment in Constraint~(\ref{eq:phi}) with Fig.~\ref{fig:eg-deployment}, where 3 OPMs are deployed in a network with 9 candidate OPM locations. In this example, $I=3$, indicates that one OPM is deployed for every 3 candidate OPM locations. Consequently, the selected candidate OPM locations to deploy OPMs are 3, 6, and 9. 
	
	These signal powers collected from OPMs will be used to achieve multi-failure localization using our proposed RINN, described in the next section.
	
	\section{Rules-Informed Neural Networks for Multi-Failure Localization}
	\label{section-rules-informed-neural-networks}
	
	In this section, we describe the pipeline of our proposed rules-informed neural network, which levearges a rules-based reasoning algorithm to decrease the size of the input fed to the ANN used for failure localization, making it more scalable and suitable to high-degree ROADM-based optical networks. Then, we analyze the corresponding computational complexity. At last, we discuss the benchmark algorithms, and our dataset generator used to train and test all approaches.
	
	\subsection{Pipeline of Proposed Solution}
	
\begin{figure}[tb]
		\centering
		\includegraphics[width=1.0\linewidth]{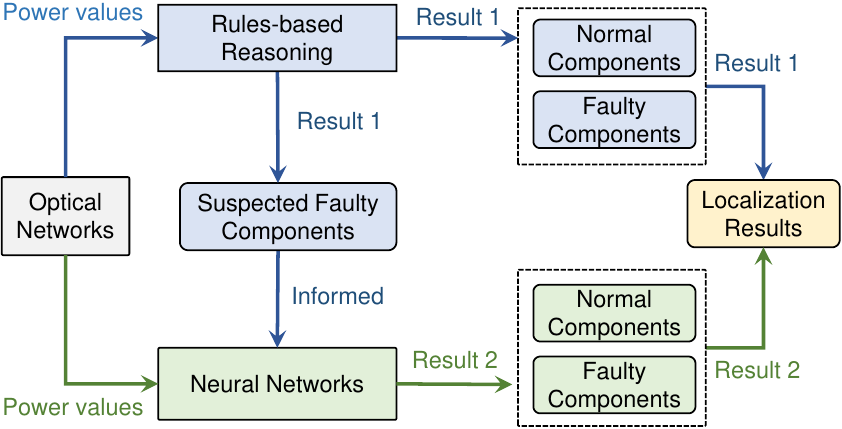}
		\caption{Pipeline of rules-informed neural networks.}
		\label{Fig-pipeline-RINN}
	\end{figure}

	The pipeline of the proposed solution consists of two stages,  namely, \textit{rules-based reasoning} marked in blue and \textit{neural networks} marked in green, as shown in Fig.~\ref{Fig-pipeline-RINN}. 
	Since the number of components in high-degree ROADM-based optical networks is high (e.g., the number of components is 5306 for the 14-node national Japan topology used in our study), it is hard to accurately localize failures for all components with only neural networks. 
	The proposed rules-based reasoning algorithm applies a threshold-based pre-classification based on the power values in each LP, and pre-classifies the components into normal components, faulty components and suspected faulty components. After this rules-based reasoning, the neural network only needs to further classify a small amount of suspected faulty components into normal components and faulty components, achieving a scalable multi-failure localization pipeline. 
	
	\subsection{Rules-based Reasoning Algorithm}
	Given $\mathcal{L}$ as the set of LPs in the network, assume that $p_l$ is the number of components in LP $l \in \mathcal{L}$. The components in all the LPs are denoted with $z$ as shown in Eqn.~\eqref{eq:LP_elements}, where $z_{l,i}$ denotes the $i$-th component in LP $l \in \mathcal{L}$. In each LP, one OPM can be deployed between any two adjacent components. The obtained power values from OPM are denoted with $x$ as in Eqn.~(\ref{eq:power_before_failure}), where $x_{l,i}$ denotes the power value in the monitoring location between the $i$-th component and the $(i+1)$-th component in LP $l \in \mathcal{L}$. If no OPM is deployed between the $i$th component and the $(i+1)$-th component in LP $l \in \mathcal{L}$, the corresponding power value $x_{l,i}$ is assigned with a small negative constant value $\alpha$ as a flag for no OPM deployed. 
	For the last component (i.e., transponder) in the LP, although the signal terminates in the transponder and no OPM can be deployed after the transponder, we can still infer if any components fail in the LP by detecting if the transponder can successfully receive the signal. Thus, if the transponder (the $p_l$-th component in LP $l$) can successfully receive the signal, $x_{l, p_l}$ is assigned with a fixed constant equal to 1 as a flag for successfully receiving the signal. Otherwise, $x_{l, p_l}$ is assigned with a fixed constant equal to 0. 
	
	\begin{equation}
		z = 
		\begin{cases}
			\left[ z_{1,1}, z_{1,2}, ..., z_{1, p_{1}} \right] \\
			\left[ z_{2,1}, z_{2,2}, ..., z_{2, p_{2}} \right] \\
			... \\
			\left[ z_{|\mathcal{L}|,1}, z_{|\mathcal{L}|,2}, ..., z_{|\mathcal{L}|,p_{|\mathcal{L}|}} \right]
		\end{cases}
		\label{eq:LP_elements}
	\end{equation}
	
	\begin{equation}
		x = 
		\begin{cases}
			\left[ x_{1,1}, x_{1,2}, ..., x_{1, p_{1}} \right] \\
			\left[ x_{2,1}, x_{2,2}, ..., x_{2, p_{2}} \right] \\
			... \\
			\left[ x_{|\mathcal{L}|,1}, x_{|\mathcal{L}|,2}, ..., x_{|\mathcal{L}|,p_{|\mathcal{L}|}} \right]
		\end{cases}
		\label{eq:power_before_failure}
	\end{equation}

	We develop a rules-based reasoning algorithm based on the power values monitored from OPMs. Specifically, we identify normal, suspected faulty and faulty components based on two key rules.
	(1) Two thresholds $\delta_{l,i}$ and $\tau_{l,i}$ $(\delta_{l,i} \leq \tau_{l,i})$ are designed to judge the status of component $z_{l,i}$.  The procedures of classifying components according to the thresholds are elaborated as follows. 
	We first calculate the optical power change $\overline{x}_{l,i}$ using the equation $\overline{x}_{l,i} = |x_{l,i} - x_{l,i-1}|$, where $x_{l,i-1}$ and $x_{l,i}$ denote the power values collected from ($i-1$)-th OPM and $i$-th OPM in lightpath $l$, respectively. The algorithm identifies the status for different components separately. For amplifiers, if the optical power change $\overline{x}_{l,i}$ is above a given threshold $\tau_{l,i}$, the component $z_{l,i}$ is normal; if the optical power change $\overline{x}_{l,i}$ is low another given threshold $\delta_{l,i}$, the component $z_{l,i}$ is faulty. For WSS and fiber spans, if the optical power change $\overline{x}_{l,i}$ is above a given threshold $\tau_{l,i}$, the component $z_{l,i}$ is faulty; if the optical power change $\overline{x}_{l,i}$ is low another given threshold $\delta_{l,i}$, the component $z_{l,i}$ is normal. Otherwise (i.e., $\delta_{l,i} \leq \overline{x}_{l,i} \leq \tau_{l,i}$), the component $z_{l,i}$ is considered as a suspected faulty component.
	(2) We design another threshold $\epsilon_{l,i}$ to judge the status of those components before OPM $i$ in lightpath $l$. If the power value $x_{l,i}$ becomes higher than a threshold $\epsilon_{l,i}$, all the components before the $i$-th OPM in LP $l$ are normal. Otherwise, at least one component before the OPM $i$ in lightpath $l$ fails.
	
	Let us now discuss how to decide the thresholds $\delta_{l,i}$ and $\tau_{l,i}$. Although each component has its component specification, e.g., the normal region of amplifier gain for amplifiers, the real-time value of components is not always equal to its component specification (with a small jitter). Due to the oscillatory nature, the thresholds should be strictly designed to ensure that the component is not misclassified.
	We first collect the last $|T|$ monitoring power change values, which is denoted with $\overline{x}_{l,i,t}$ $(t \in T)$. We obtain the status of component $z_{l,i}$ for $\overline{x}_{l,i,t}$, where the binary variable $\overline{\gamma}_{l,i,t} = 1$ denotes normal, while $\overline{\gamma}_{l,i,t} = 0$ denotes faulty. 
	We divide $\overline{x}_{l,i,t}$ $(t \in T)$ into two sets: set $P$ is defined as the index of the monitoring power change of normal component, which can be obtained by setting $P = \{p | p \in N: \overline{\gamma}_{l,i,p} = 1\}$. Similarly, set $Q$ is defined as the index of the monitoring power change of the faulty component, which can be obtained by setting $Q = \{q | q \in N: \overline{\gamma}_{l,i,q} = 0\}$.    
	If component $z_{l,i}$ is an amplifier, the thresholds can be calculated using the equations $\delta_{l,i} = ave \{\overline{x}_{l,i,t} | t \in Q, \overline{x}_{l,i,t} < \hat{M} \}$ and $\tau_{l,i} = ave \{\overline{x}_{l,i,t} | t \in P, \overline{x}_{l,i,t} > \overline{M} \}$, where $\hat{M} = min\{\overline{x}_{l,i,t} | n \in P\}$ and $\overline{M} = max \{\overline{x}_{l,i,t} | t \in Q\}$. 
	If component $z_{l,i}$ is not an amplifier, we obtain the thresholds by the equations $\tau_{l,i} = ave \{\overline{x}_{l,i,t} | t \in Q, \overline{x}_{l,i,t} > \hat{m}\}$ and $\delta_{l,i} = ave \{\overline{x}_{l,i,t} | t \in P, \overline{x}_{l,i,t} < \overline{m} \}$, where $\hat{m} = max \{\overline{x}_{l,i,t} | t \in P\}$ and $\overline{m} = min \{\overline{x}_{l,i,t} | t \in Q\}$.
	Similarly, we decide the threshold $\epsilon_{l,i}$ using last $|T|$ power values $x_{l,i,t}$ $(t \in T)$ collected from $i$-th OPM in LP $l$. Specifically, the threshold $\epsilon_{l,i}$ is calculated using the equation $\epsilon_{l,i} = ave \{x_{l,i,t} | t \in P, x_{l,i,t} > \overline{M}\}$, where $\overline{M} = max \{x_{l,i,t} | n \in Q \}$.
	There are more new power values with the running of optical networks. The existing thresholds may be not suitable for the new status. Therefore, we will repeat to collect power values, and then update the thresholds using the newest $|T|$ values. The above method is named as moving-window-based threshold determination.

	\begin{algorithm}[tb]
		\caption{Rules-based Reasoning Algorithm}
		\small
		\label{alg:algorithm1}
		\DontPrintSemicolon
		\KwIn{$\mathcal{L}, z, x$, $\delta$, $\tau$, $\epsilon$, $\alpha$} 
	\KwOut{$N_{normal}$, $N_{failure}$, $N_{suspect}$}
	Initialize $N_{all} \leftarrow \{\}$, $N_{normal} \leftarrow \{\}$, $N_{failure} \leftarrow \{\}$, $N_{suspect} = \{\}$ \\
	\For{each LP $l \in \mathcal{L}$}
	{
		$i_s \leftarrow 0$ \\
		\For{$i=1$ to $p_l$}
		{
			$N_{all} \leftarrow N_{all} \cup \{ z_{l,i} \}$ \\
			\If{$i \geq 2$ $\land$ $i < p_l$ $\land$ $x_{l,i-1} \neq \alpha$ $\land$ $x_{l,i} \neq \alpha$}
			{
				$\overline{x}_{l,i} \leftarrow |x_{l,i} - x_{l,i-1}|$\\
				
				\If{$z_{l,i}$ \text{is an amplifier}} 
				{
					\If{$\overline{x}_{l,i} \geq \tau_{l,i}$}
					{
						$N_{normal} \leftarrow N_{normal} \cup \{ z_{l,i} \}$
					}
					\If{$\overline{x}_{l,i} < \delta_{l,i}$} 
					{
						$N_{failure} \leftarrow N_{failure} \cup \{ z_{l,i} \}$ 
					}
				}
				
				\If{$z_{l,i}$ is not an amplifier} 
				{
					\If{$\overline{x}_{l,i} \geq \tau_{l,i}$} 
					{
						$N_{failure} \leftarrow N_{failure} \cup \{ z_{l,i} \}$
					}
					\If{$\overline{x}_{l,i} < \delta_{l,i}$} 
					{
						$N_{normal} \leftarrow N_{normal} \cup \{ z_{l,i} \}$
					}
				}
			}
			\If{$x_{l,i} \geq \epsilon_{l,i}$} 
			{
				$i_s \leftarrow i$
			}
		}
		
		\If{$i_s > 0$} 
		{
			\For{$j=1$ to $i_s$}
			{
				$N_{normal} \leftarrow N_{normal} \cup \{ z_{l,j} \}$
			}
		}
	}
	$N_{suspect} \leftarrow N_{all} \setminus (N_{normal} \cup N_{failure}$) \\
	Return $N_{normal}$, $N_{failure}$, $N_{suspect}$
\end{algorithm}

The proposed rules-based reasoning algorithm is shown in Algorithm~\ref{alg:algorithm1}.
The inputs are the set of LPs $\mathcal{L}$, the components $z$, the power values $x$, the thresholds $\delta$, $\tau$ and $\epsilon$ and a negative constant value $\alpha$.
The outputs are the set of normal components, the set of faulty components, and the set of suspected faulty components. The algorithm starts by initializing four empty sets for all traversed components, normal components, faulty components, and suspected faulty components, which are denoted by $N_{all}$, $N_{normal}$, $N_{failure}$, $N_{suspect}$, respectively (line 1). 
Then, the algorithm loops over all LPs to check the corresponding power values in the OPMs of each LPs and determine the status of all components (line 2-35).
During the loop over OPMs in each LP, the algorithm sets $i_s$ to 0 (line 3), and then identifies the status of components between two adjacent OPMs.
The algorithm first determines the status of components based on optical power change $\overline{x}_{l,i}$ in adjacent OPMs (line 6-24).
We calculate the optical power change $\overline{x}_{l,i}$ passing through component $z_{l,i}$ in line 7, and then judge which the component $z_{l,i}$ is an amplifier or a WSS/fiber in line 8 and 16, respectively.
If the component $z_{l,i}$ is an amplifier, we compare $\overline{x}_{l,i}$ with the thresholds $\delta_{l,i}$ and $\tau_{l,i}$ in line 9 and 12, and then judge the status of component $z_{l,i}$ in line 10 and 13.
If the component $z_{l,i}$ is a WSS or fiber span (line 16), we also compare $\overline{x}_{l,i}$ with the thresholds $\delta_{l,i}$ and $\tau_{l,i}$ in line 17 and 20, and then identify the status of component $z_{l,i}$ in line 18 and 21.
Then, the algorithm continues to determine the status of components based on normal OPMs. Specifically, 
the algorithm compares the threshold $\epsilon_{l,i}$ with the monitor value $x_{l,i}$ in line 25. If the monitor value $x_{l,i}$ is above the threshold $\epsilon_{l,i}$, we set all the components before OPM $i$ to be normal components (line 29-34).
After that, the set of suspected faulty components is obtained by removing all the determined normal components and faulty components from the set of all traversed components (line 35).

\begin{figure}[tb]
	\centering
	\includegraphics[width=1\linewidth]{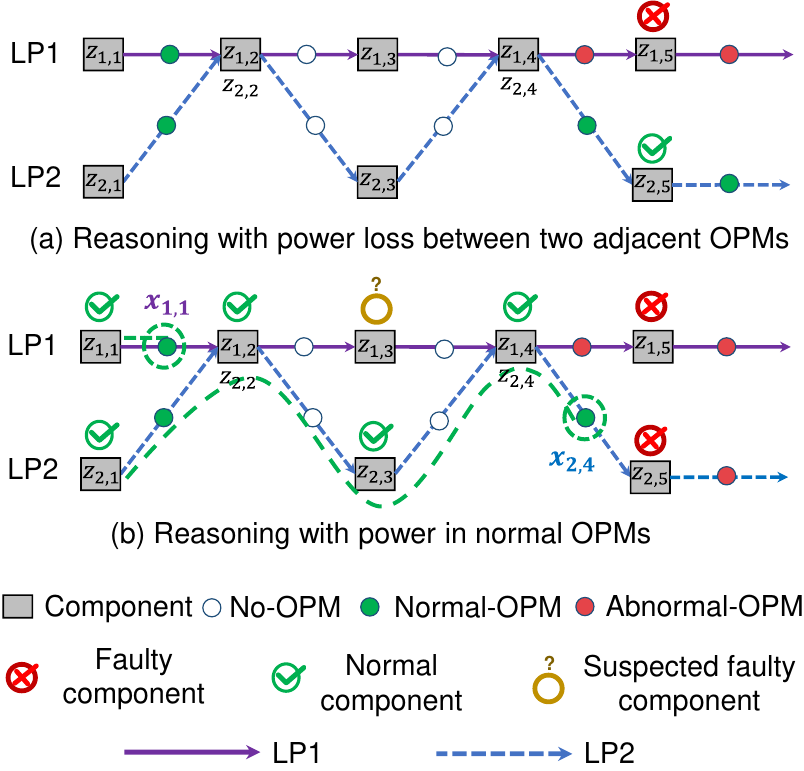}
	\caption{Illustration of rules-based reasoning algorithm.}
	\label{Fig-Rules-based-Reasoning}
\end{figure}

The proposed rules-based reasoning algorithm is illustrated with an example in Fig.~\ref{Fig-Rules-based-Reasoning}, representing two lightpaths LP1 (in purple solid line) and LP2 (in blue dotted line). 
First, the algorithm observes the optical power change between two adjacent OPMs as shown in Fig.~\ref{Fig-Rules-based-Reasoning}(a). 
We define adjacent OPMs only those OPM that are located both before and after a component. In this example, the adjacent OPMs are only the OPM before and after component $z_{1,5}$, and the OPM before and after component $z_{2,5}$. 
After comparing the optical power change between these adjacent OPMs, component $z_{1,5}$ and component $z_{2,5}$ are determined to be faulty and normal, respectively.
The algorithm then further identifies normal components, as shown in Fig.~\ref{Fig-Rules-based-Reasoning}(b). In LP 1, since the monitor value $x_{1,1}$ between component $z_{1,1}$ and component $z_{1,2}$ is above the threshold $\epsilon_{1,1}$, component $z_{1,1}$ is determined to be normal. For LP 2, since the monitor value $x_{2,4}$ between components $z_{2,4}$ and $z_{2,5}$ is above the threshold $\epsilon_{2,4}$, all the components traversed in the LP 2 before this OPM (i.e., components $z_{2,1}$, $z_{2,2}$, $z_{2,3}$ and $z_{2,4}$) are normal.

\textbf{Complexity analysis}: The complexity of the rules-based reasoning algorithm is analyzed as follows. Assume that the maximum number of hops of LP is $O(p_{max})$. Then, for each LP, the complexity of reasoning with power loss between two adjacent OPMs (line 4-19) is $O(p_{max})$. Moreover, the complexity of reasoning with power in normal OPM (line 20-25) is also $O(p_{max})$. Hence, the overall complexity of the rules-based reasoning algorithm for all the LPs is $O(p_{max}|\mathcal{L}|)$. 

\subsection{Rules-Informed Neural Networks for Multi-Failure Localization}
\label{sec:subsec_neural_network}
An artificial neural network (ANN) model is designed to further predict the status (i.e., failure or no-failure) of all the suspected faulty components. 
The model takes the power values of each suspected faulty component (as identified by the rules-based reasoning algorithm) and outputs the status of the component (normal or faulty).
Therefore, the problem of assessing the status of all suspected faulty components is modeled as a set of binary classification problems to classify all the components. 
Note that we avoid using a neural network to simultaneously predict the status of total suspected faulty components because the number of suspected faulty components varies with each failure scenario. To generalize ANN across different failure scenarios, padding would be required to account for all suspected faulty components. 

\begin{figure}[htb]
	\centering
	\includegraphics[width=1.0\linewidth]{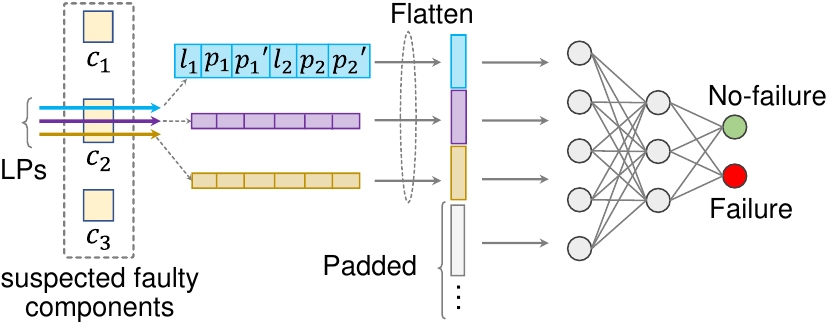}
	\caption{Illustration of failure localization with RINN.} 
\label{Fig-neural-networks}
\end{figure}

The proposed ANN model will analyze the power values of OPMs for each suspected faulty component from the rules-based reasoning algorithm, as shown in Fig.~\ref{Fig-neural-networks}. 
Specifically, for each suspected faulty component (e.g., component $c_2$ in the example in the figure), the inputs contain the information related to power values from OPMs of all the LPs traversing the component. 
The information used for each LP traversing the component includes the number of hops in the preceding OPM before the suspected faulty component (defined as $l_1$), the corresponding power values in the last monitoring time (defined as $p_1$), and the power values in the current monitoring time (defined as $p_{1}'$).
In addition, the inputs also contain the number of hops in the subsequent OPM after the suspected faulty component (defined as $l_2$), the corresponding power values in the last monitoring time (defined as $p_2$), and the power values in the current monitoring time (defined as $p_2'$). 
If no component exists before or after the component in the LP, the corresponding inputs are set to 0. 
All information is flattened into a vector, whose length matches the maximum number of LPs traversing any component, with the upper bound being the total number of LPs. 
For components with fewer traversing LPs, the remaining slots in the input vector are padded with 0 to maintain a uniform length. 
This work adopts Binary Cross-Entropy Loss as shown in Eqn.~(\ref{equ-cross-entry}), where $L$ denotes the loss value, $\sigma(x) = 1 / \left(1 + e^{-x}\right)$, $y$ denotes if component $c$ is faulty, and $\hat{y}$ denotes the predicted result of the current component, respectively. 
\vspace{-3pt}
\begin{equation}
\begin{aligned}
	L = - \left[ y \cdot \log \left(\sigma(\hat{y} \right) + \left(1 - y\right) \cdot \log \left(1 - \sigma(\hat{y} \right) \right]
\end{aligned}
\label{equ-cross-entry}
\end{equation}

\subsection{Benchmark Methods}
We compare our proposed RINN to two baselines, namely, rules-based reasoning and artificial neural networks, to evaluate the advantages of integrating rules-based reasoning and artificial neural networks. Note that all the three approaches above adopt same input features, i.e., the power values collected by OPMs.

\subsubsection{Rules-based benchmark (Rules)}
The rules-based reasoning can not always directly find all the faulty components, as some components may only be classified into the set of suspected faulty components. If the set of suspected faulty components is not an empty, we randomly select some components from it with equal probability. Then the union of the faulty set from the rules-based reasoning and the randomly selected components from the suspected faulty components are considered as the determined faulty components for the \textit{Rules} baseline.

\subsubsection{Artificial-neural-networks-based benchmark (ANN)}
We take the ANN model~\cite{mayer2021machine} as a benchmark, and it follows the architecture introduced in Sec.~\ref{sec:subsec_neural_network}. 
Unlike the proposed RINN, which incorporates localization results from rules-based reasoning about suspected faulty components, the ANN benchmarks lack this context and, therefore, are applied to all components to identify faulty components.

\begin{algorithm}[htb]
\caption{Dataset Generator}
\small
\label{alg:algorithm_generator}
\DontPrintSemicolon
\KwIn{Network topology, Component parameters,
	Number of LP requests ($|\mathcal{L}|$), Number of samples ($S$), Set of possible number of failures ($N_{f}$), Number of placed OPM $M'$
}
\KwOut{Training samples}

\For{$l \gets 1$ \KwTo $|\mathcal{L}|$}{ 
	Randomly generate a source/destination pair\;
	Route, and assign fiber and wavelength with SPFF algorithm considering wavelength continuity\;
}

\For{each LP $l \leftarrow 1$ \KwTo $|\mathcal{L}|$} { 
	\If{each candidate OPM location in LP $l$} {
		\If{OPM is deployed in the current candidate OPM location} {
			Obtain the monitoring values before failures 
		}
	}
}

\For{$s \gets 1$ \KwTo $S$}{
	Randomly select one number $n_f$ from $N_{f}$\; 
	\For{$c \gets 1$ \KwTo $n_{f}$}{
		Determine the failure location, failure type, and failure effect
	}
	\For{each LP $l \leftarrow 1$ \KwTo $|\mathcal{L}|$} { 
		\If{each candidate OPM location in LP $l$} {
			\If{OPM is deployed in the current candidate OPM location} {
				Obtain the monitoring values after failures\; 
			}
		}
	}
}

\end{algorithm}

\subsection{Dataset Generator}

We develop a simulator to generate the dataset as shown in Algorithm~\ref{alg:algorithm_generator}. 
The inputs of the algorithm include the network topology, parameters of components (note that these parameters will be described in Sec.~\ref{section-experimental-setup-and-results}), number of LP requests, number of samples, set of possible numbers of failures, and the number of placed OPM. The outputs are the values of power for all the signals at each OPM in the network, which compose our dataset. 
The Algorithm~\ref{alg:algorithm_generator} first generates $|\mathcal{L}|$ source/destination pair and then serves the requests with the shortest path and first fit (SPFF) algorithm to finish the route, fiber, and wavelength assignment (RFWA). 
After obtaining LP information, Algorithm~\ref{alg:algorithm_generator} starts to obtain the normal OPM values before the failure (line 5-10) and the abnormal OPM values after failure (line 17-23). Regarding the normal OPM values before the failure, the algorithm loops over each LP (line 5) and obtains power values from each deployed OPM (line 6-10). Then, Algorithm~\ref{alg:algorithm_generator} generates the OPM values after failure for each possible failure sample $s \in S$.
Specifically, Algorithm~\ref{alg:algorithm_generator} first randomly selects the number of failed components $n_f$ from $N_f$ (line 13) and decides the corresponding failure locations, failure types, and failure effects for all failures (line 14-16). After obtaining the information about failures, Algorithm~\ref{alg:algorithm_generator} obtains the monitoring values after failures for all the LPs (line 17-23). 

\begin{figure}[htb]
\centering
\includegraphics[width=1.0\linewidth]{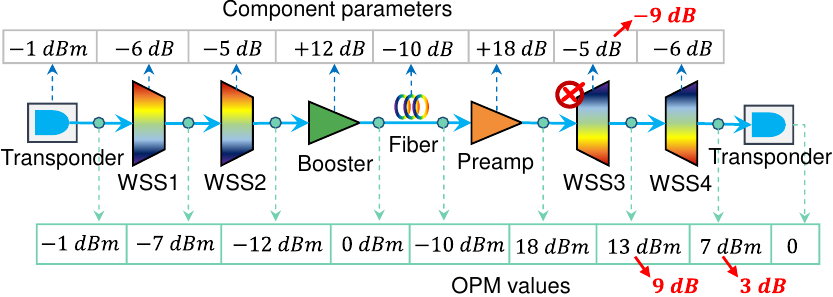}
\caption{Illustration of dataset generator.}
\label{Fig-dataset-generator}
\end{figure}

Fig. \ref{Fig-dataset-generator} shows an example of the dataset generator with an LP. The top table is the component parameters, where \lq\lq$-$'' denotes attenuation, and \lq\lq$+$\rq\rq denotes amplifier. We also assume the failure location is WSS3 with an extra insertion loss of $4$ $dB$. Thus, each OPM value can be calculated according to component parameters and failure effect, as shown in the bottom table.

\section{Illustrative Numerical Results}
\label{section-experimental-setup-and-results}

In this section, we first discuss the simulation setup and benchmark methods. Then, we compare the performance of our proposed RINN to the benchmark methods. Finally, we perform experimental demonstrations to validate the proposed approach in a real testbed. 

\begin{figure}[tb]
\centering
\includegraphics[width=1.0\linewidth]{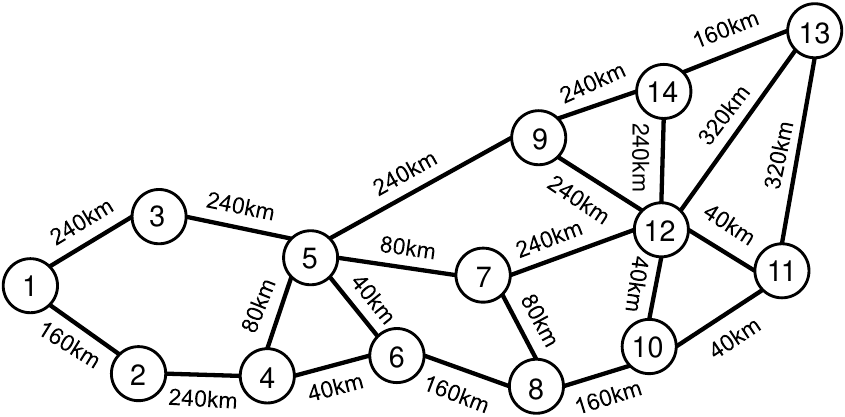}
\caption{Simulation topology.}
\label{Fig-japnet}
\end{figure}

\subsection{Simulation Setup}
We perform the simulations with a custom-built Python and Pytorch simulator on a commercial server (CPU with i9-10980XE and RAM with 32G). 
We evaluate the performance of our proposed RINN on Japan topology with a fiber span of 80 kilometers~\cite{9355394}, as shown in Fig.~\ref{Fig-japnet}. 
We utilize $1 \times 32$ local-WSS~\cite{huawei-P32-WSS} with insertion loss between 3.3 dB to 6.8 dB~\cite{8386355} together with $8 \times 24$ line-WSS~\cite{8386355} with insertion loss equal to 5 dB~\cite{10201922}. 
Preamp gain, Booster gain, and ILA gain are between 18 and 32 dB, 10 and 20 dB, and 20 and 32 dB, respectively~\cite{ibrahimi2020minimum}. 
All the parameters for simulation are listed in Tab.~\ref{tab:parameters_simulation}. 

\begin{table}[htbp] 
\small
\caption{Parameters for simulation.}
\centering
\label{tab:parameters_simulation}
\begin{tabular}{ll} 
	\toprule 
	Params & Description\\
	\midrule
	Number of ports in local-WSS $(k)$ & $32$  \\
	Number of ports in line-WSS $(m, n)$ & $8, 24$  \\
	Transponder power & $-1$ $dBm$ \\ 
	Fiber attenuation & $0.2$ $dB/km$\\
	Insertion loss of local-WSS &  $3.3 - 6.8$ $dB$  \\
	Insertion loss of line-WSS &  $5$ $dB$ \\
	Preamp gain &  $18 - 32$ $dB$  \\
	Booster gain &  $10 - 20$ $dB$  \\
	ILA gain &  $20 - 32$ $dB$  \\
	Length of single fiber span ($L^{span}$) & $80$ $km$ \\
	Set of possible number of failures ($N_{f}$) & $[1], [2], [3], [1, 2, 3]$ \\
	Number of training and testing samples & 1000, 1000 \\
	Learning rate & 0.0001 \\
	Number of training epochs & 100 \\
	Number of training samples & 2000 \\
	Number of test samples & 1000 \\
	\bottomrule 
\end{tabular} 
\end{table}

We implement a two-layer ANN model, including input layer, single-hidden layer and output layer. The number of neurons in the input layer is determined by the maximum number of LPs traversing all components, multiplied by the six features per LP as detailed in ~\ref{sec:subsec_neural_network} and Fig. \ref{Fig-neural-networks}. The number of neurons in the hidden layer and output layer are 64, and 2, respectively. The activation function of hidden layer and output layer are sigmoid function and softmax function, respectively. The Adam algorithm is applied as training algorithm without any normalization way. The number of training epochs is 100, with learning rate equal to 0.0001. The number of training samples is 2000. The model structure and parameters used in the \textit{RINN} are consistent with those in the benchmarks.

\subsection{Performance metrics}
We evaluate the localization accuracy of the proposed RINN using three different accuracy metrics. The localization accuracy is defined as the percentages of testing samples where the failed components are successfully identified among all the testing samples. Since we consider the multi-failure scenario, the proposed RINN might identify only part of the failed components in some testing samples. 
We define two concepts, namely, \textit{complete localization accuracy} and \textit{partial localization accuracy} to evaluate the localization accuracy. Specifically, complete localization accuracy is defined as the percentage of testing samples where the faulty components identified by algorithms exactly match all the actual faulty components. Partial localization accuracy is defined as the percentage of testing samples where the algorithms identify only a partial subset of the actual faulty components (not total actual faulty components).
Moreover, we also define \textit{total localization accuracy} as the percentage of testing samples where at least one failed component can be identified\footnote{If each testing sample only contains one failed component, we consider the partial localization accuracy as \lq\lq 0\%". In this condition, complete localization accuracy equals to total localization accuracy.}. 
This encompasses both situations: when the algorithm identifies all the faulty components (complete localization accuracy) and when it only identifies some of them (partial localization accuracy). 
Consequently, the value of total localization accuracy equals the value of complete localization accuracy, plus partial localization accuracy. 

We compare the rules-informed neural networks (named as \textit{RINN}) to rules-based reasoning (named as \textit{Rules}) and artificial neural networks (named as \textit{ANN}). 
The complete localization accuracy and partial localization accuracy of these approaches are denoted with \textit{approaches (Complete)} and \textit{approaches (Partial)}, respectively. Note that we define partial localization as a performance metric to evaluate localization accuracy, while the concept of partial telemetry \cite{mayer2021machine} usually refers to the fact that only part of the network monitoring data is available.

\subsection{Investigated Failure Scenarios}
We evaluate four \textit{failure scenarios} with one failure (named as \textit{1-Failure}), two failures (named as \textit{2-Failure}), three failures (named as \textit{3-Failure}), and mixed failures with the number of failures randomly selected from one, two, and three (named as \textit{Mixed-Failure}), respectively. We then evaluate the localization accuracy under different percentage of OPMs, different types of failure, and different number of LPs, respectively.
The varying percentages of OPMs simulate different practical networks, while the number of failures, the type of failures and the number of LPs could be changing in time. Therefore, it is sufficient to generate a training dataset for each specific OPM percentage and train a separate ANN model for each one.
In the test procedure, we select the trained ANN models for testing different scenarios based solely on the percentage of OPMs rather than other factors. Note that we use some ways to avoid data leakage that the proposed \textit{RINN} learns any specific failure statistics leaked from training dataset to test dataset, which can artificially improve its performance.
Firstly, the training/test datasets have different LPs, including different source/destination pairs and routing paths.
Secondly, the samples in training/test datasets have different failure locations and failure effects (e.g., insertion loss values of WSS).
Therefore, for any given a training dataset and a test dataset, they all show the significant distinctions.
For all the evaluations, we obtain the localization accuracy from 1000 failure cases (i.e., testing samples) for each failure scenario (the number of failed components in each failure case is determined according to the failure scenario). 

\begin{figure}[tb]
\centering
\begin{subfigure}[b]{0.43\textwidth}
	\centering
	\includegraphics[width=\textwidth]{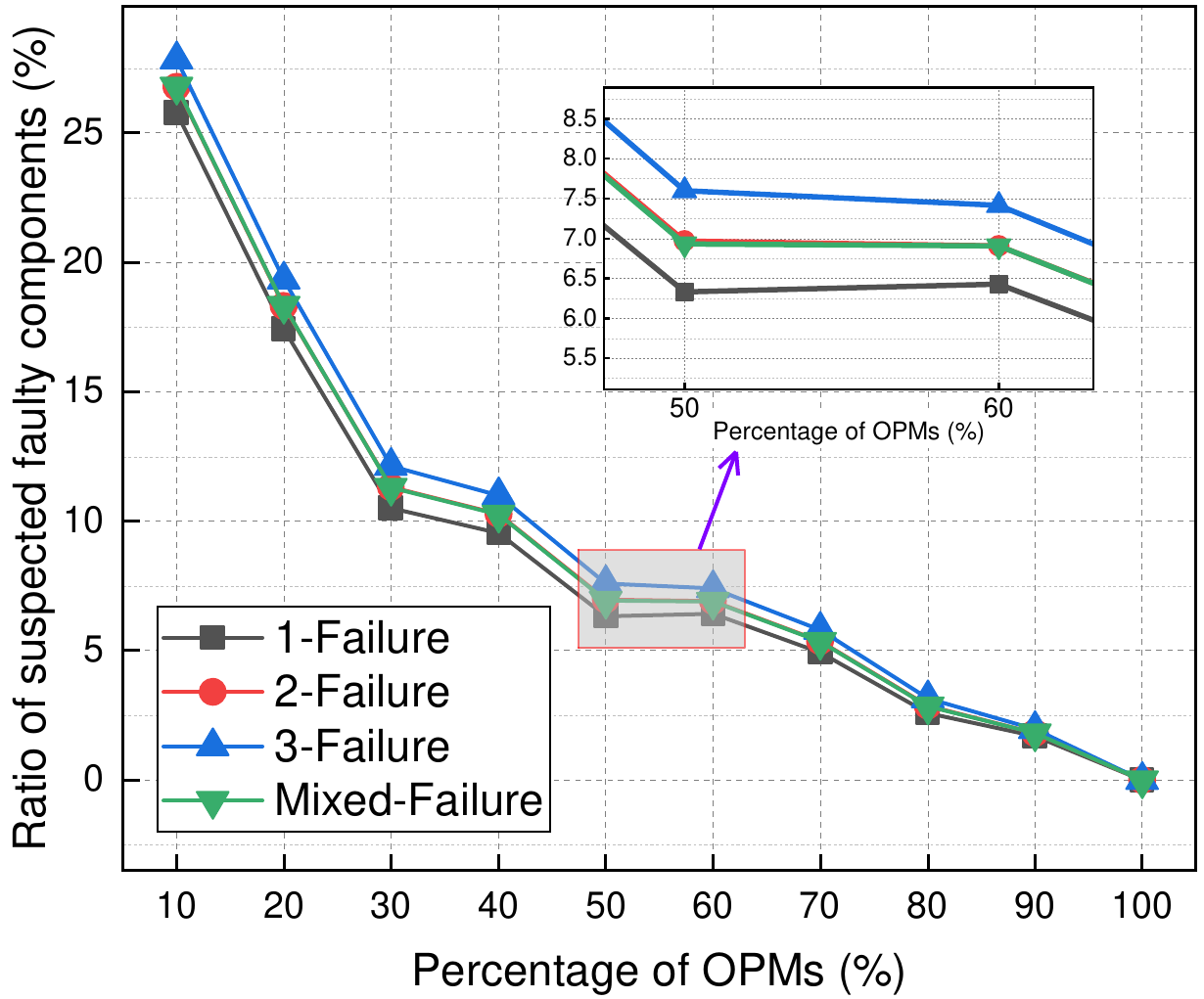}
	\caption{\centering \small Ratio of suspected faulty components vs. percentage of OPMs.}
	\label{fig:subfig_rule_OPM}
\end{subfigure}
\hfill
\begin{subfigure}[b]{0.43\textwidth}
	\centering
	\includegraphics[width=\textwidth]{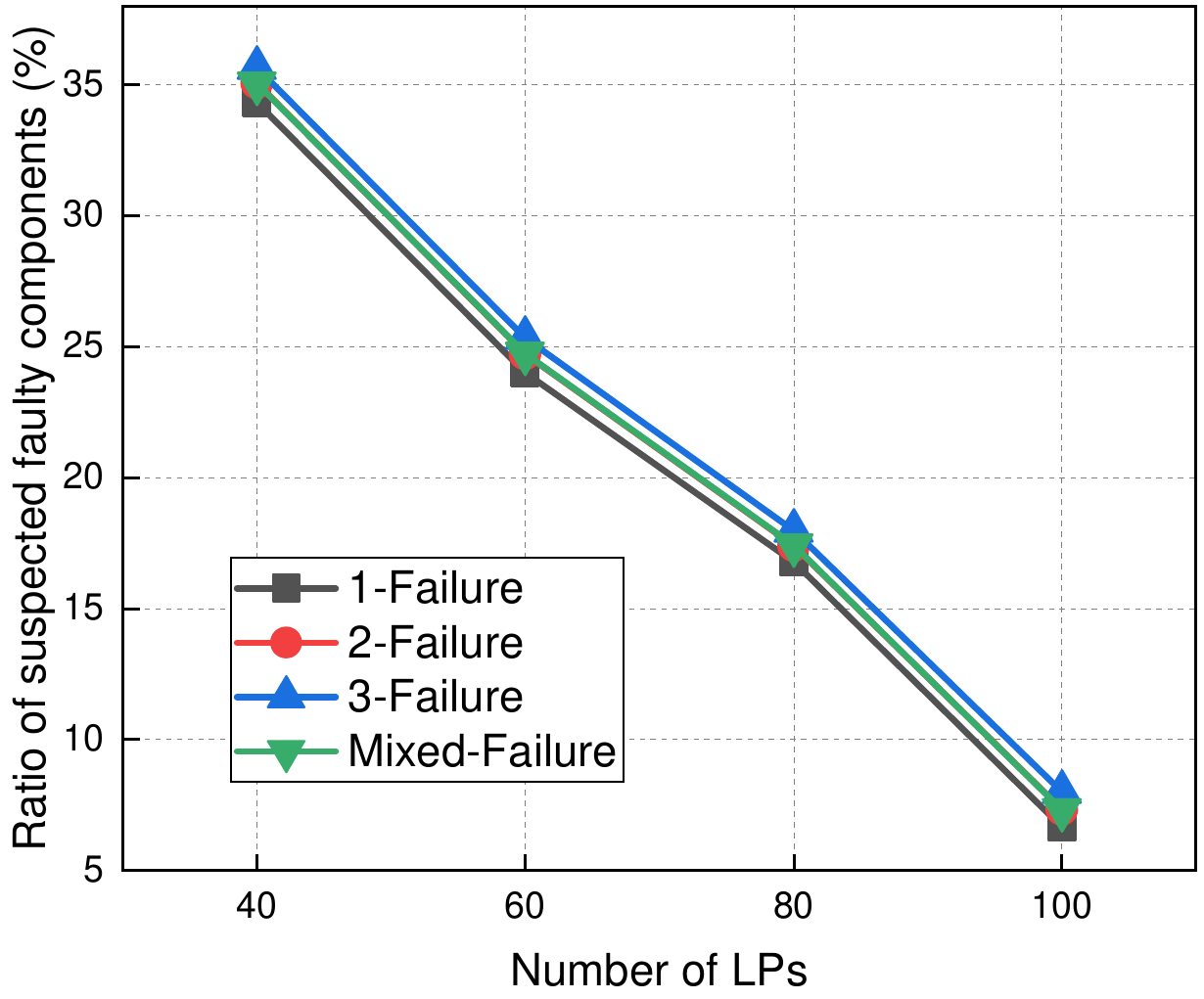}
	\caption{\centering \small Ratio of suspected faulty components vs. number of LPs.}
	\label{fig:subfig_rule_lp}
\end{subfigure}
\caption{Ratio of suspected faulty components after rules-based reasoning.}
\label{fig:evaluation_rules_based_reasoning}
\end{figure}

\begin{figure*}[htbp]
\centering
\begin{subfigure}[b]{0.24\textwidth}
	\centering
	\includegraphics[width=\textwidth]{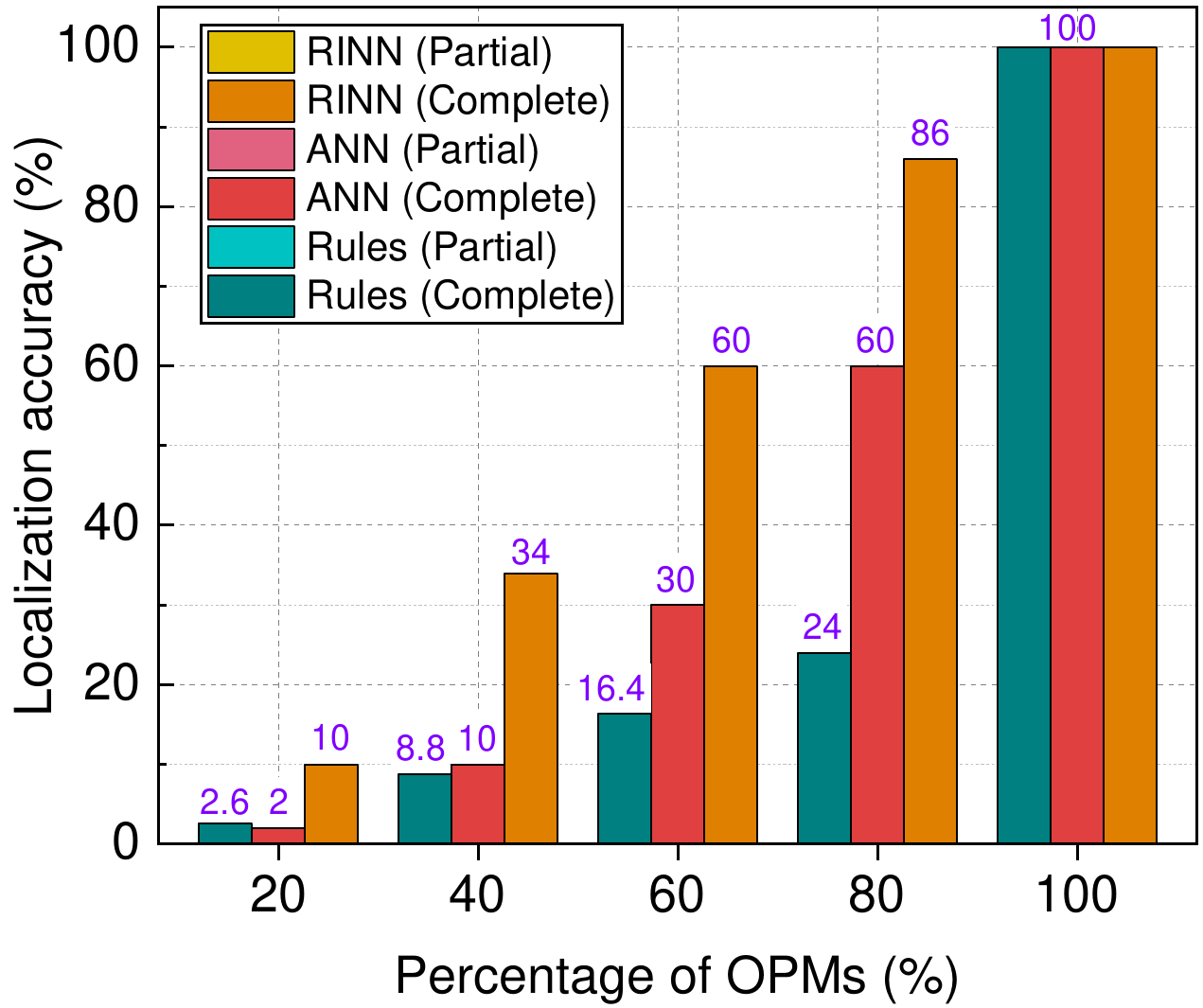}
	\caption{\centering \small 1-failure}
\end{subfigure}
\begin{subfigure}[b]{0.24\textwidth}
	\centering
	\includegraphics[width=\textwidth]{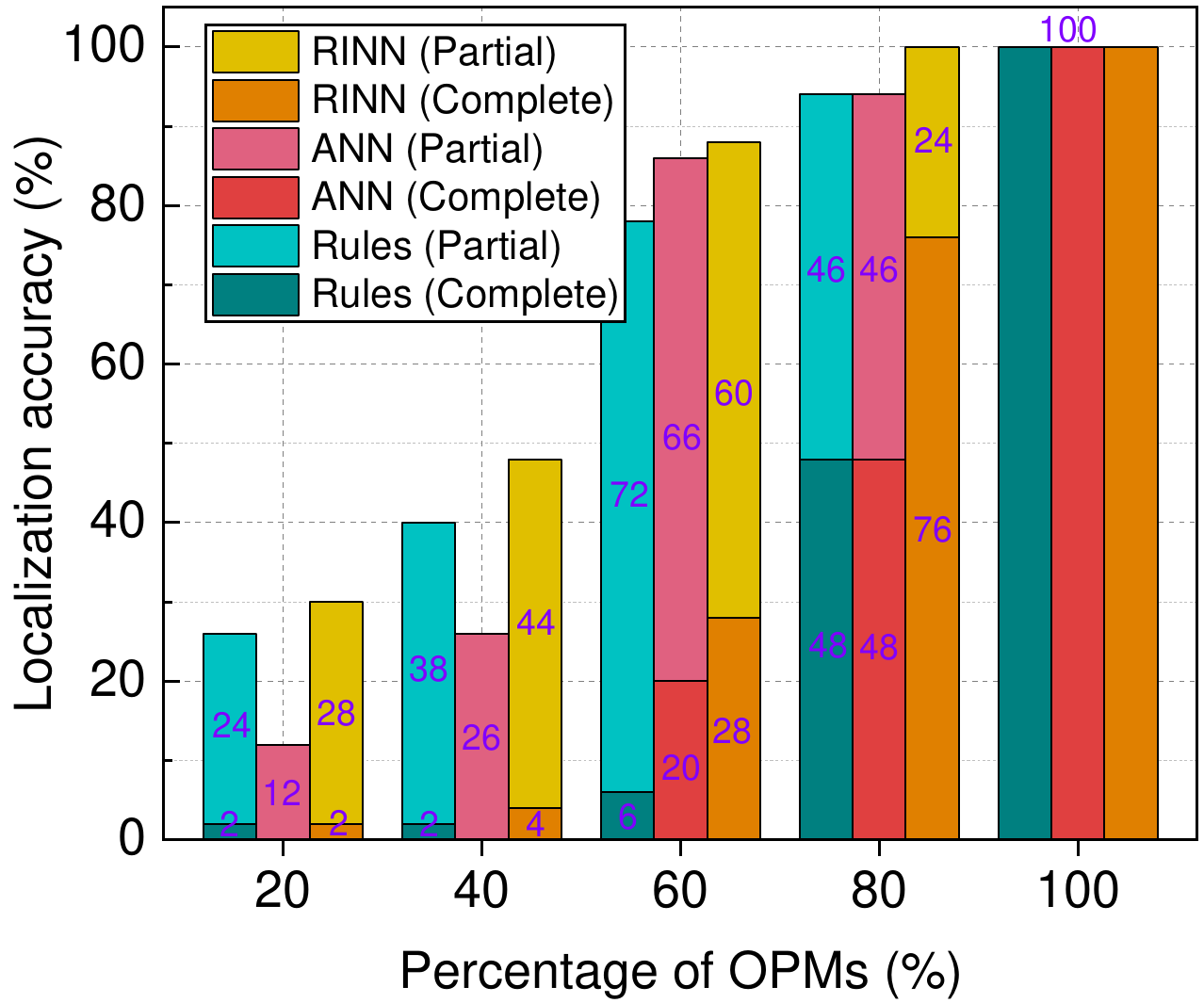}
	\caption{\centering \small 2-failure}
\end{subfigure}
\begin{subfigure}[b]{0.24\textwidth}
	\centering
	\includegraphics[width=\textwidth]{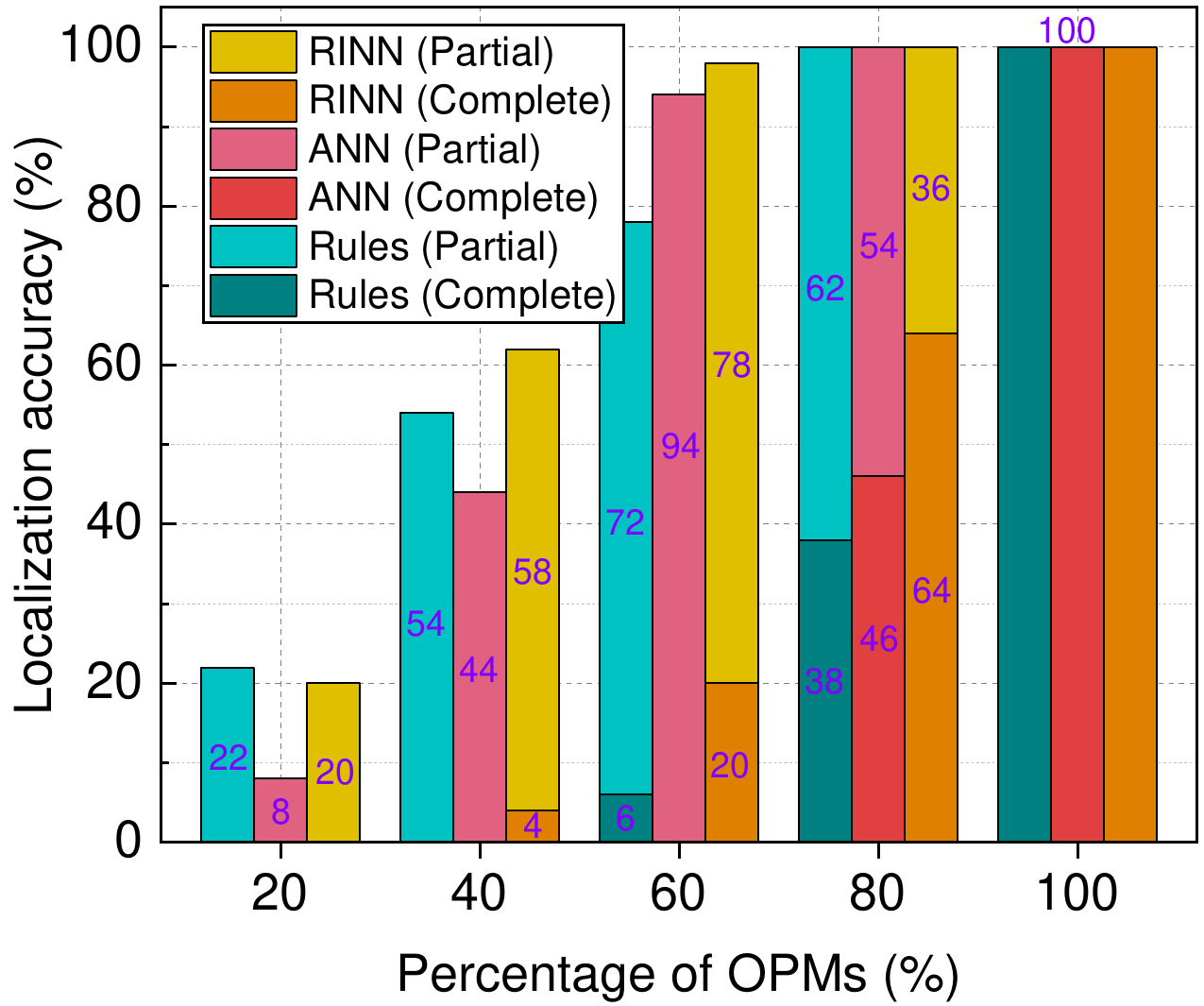}
	\caption{\centering \small 3-failure}
\end{subfigure}
\begin{subfigure}[b]{0.24\textwidth}
	\centering
	\includegraphics[width=\textwidth]{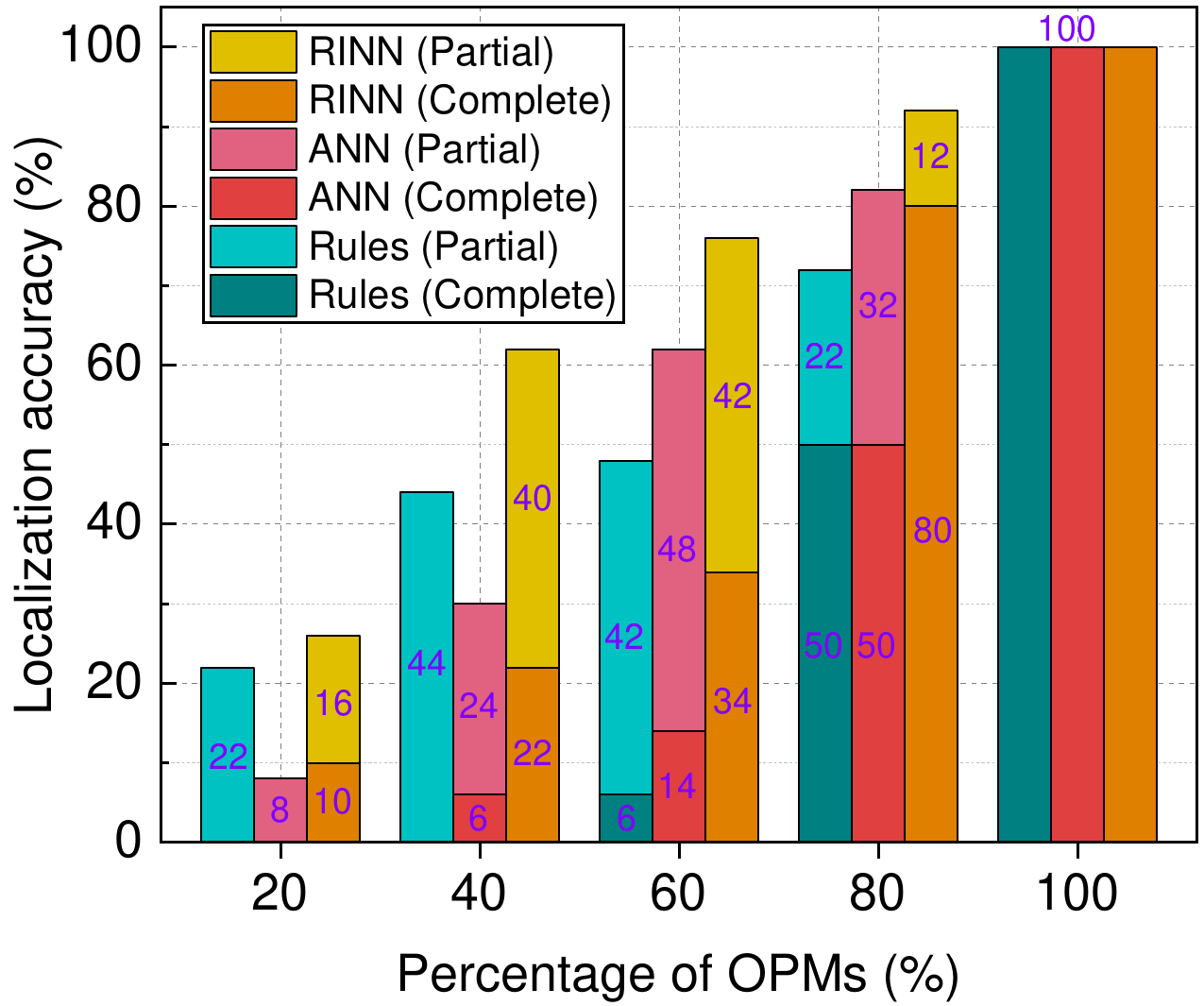}
	\caption{\centering \small mixed-failure}
\end{subfigure}
\caption{Localization accuracy under different percentages of OPMs.}
\label{fig:result_percentage_of_opm}
\end{figure*}

\begin{figure*}[htbp]
\centering
\begin{subfigure}[b]{0.24\textwidth}
	\centering
	\includegraphics[width=\textwidth]{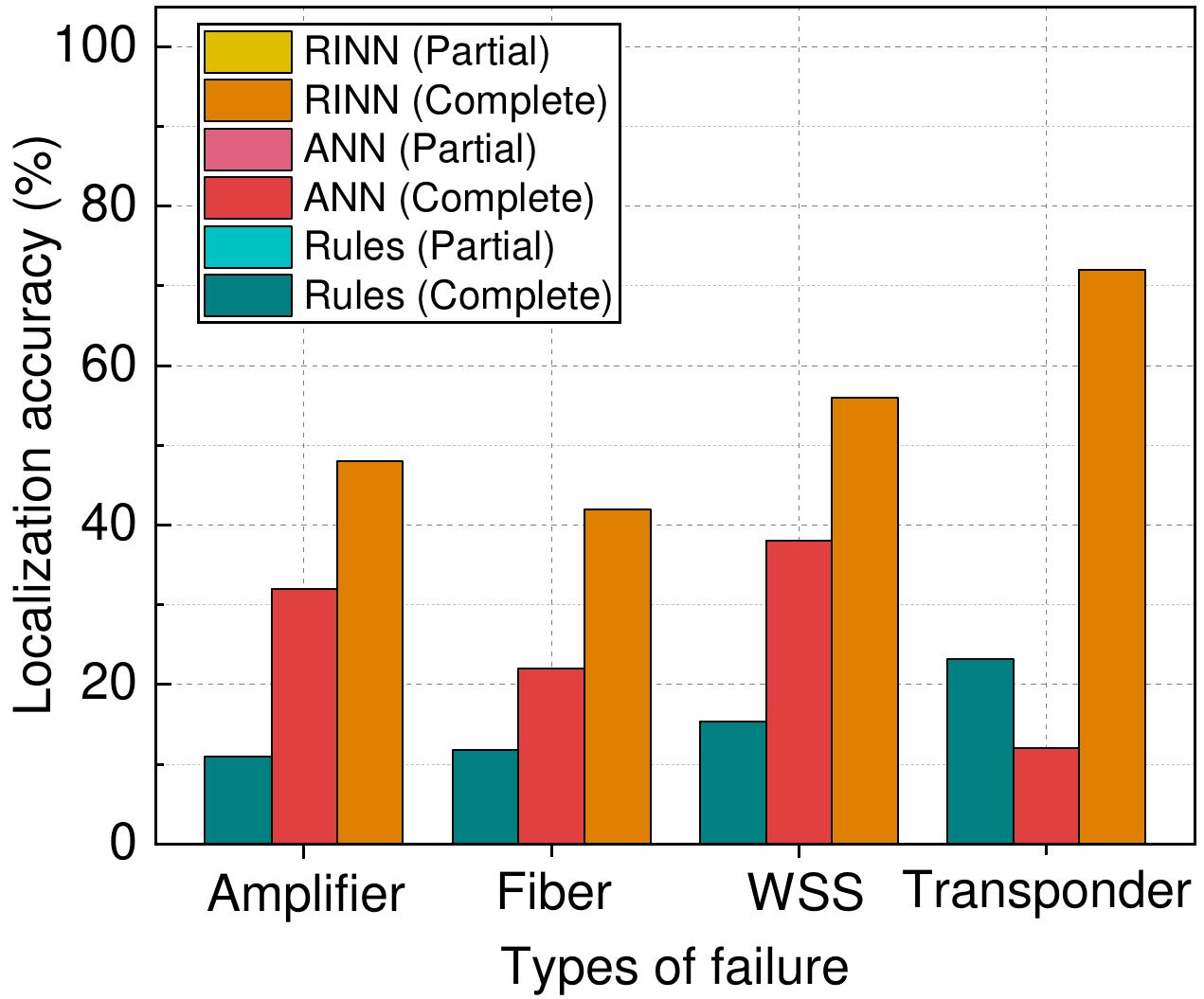}
	\caption{\centering \small 1-failure}
\end{subfigure}
\begin{subfigure}[b]{0.24\textwidth}
	\centering
	\includegraphics[width=\textwidth]{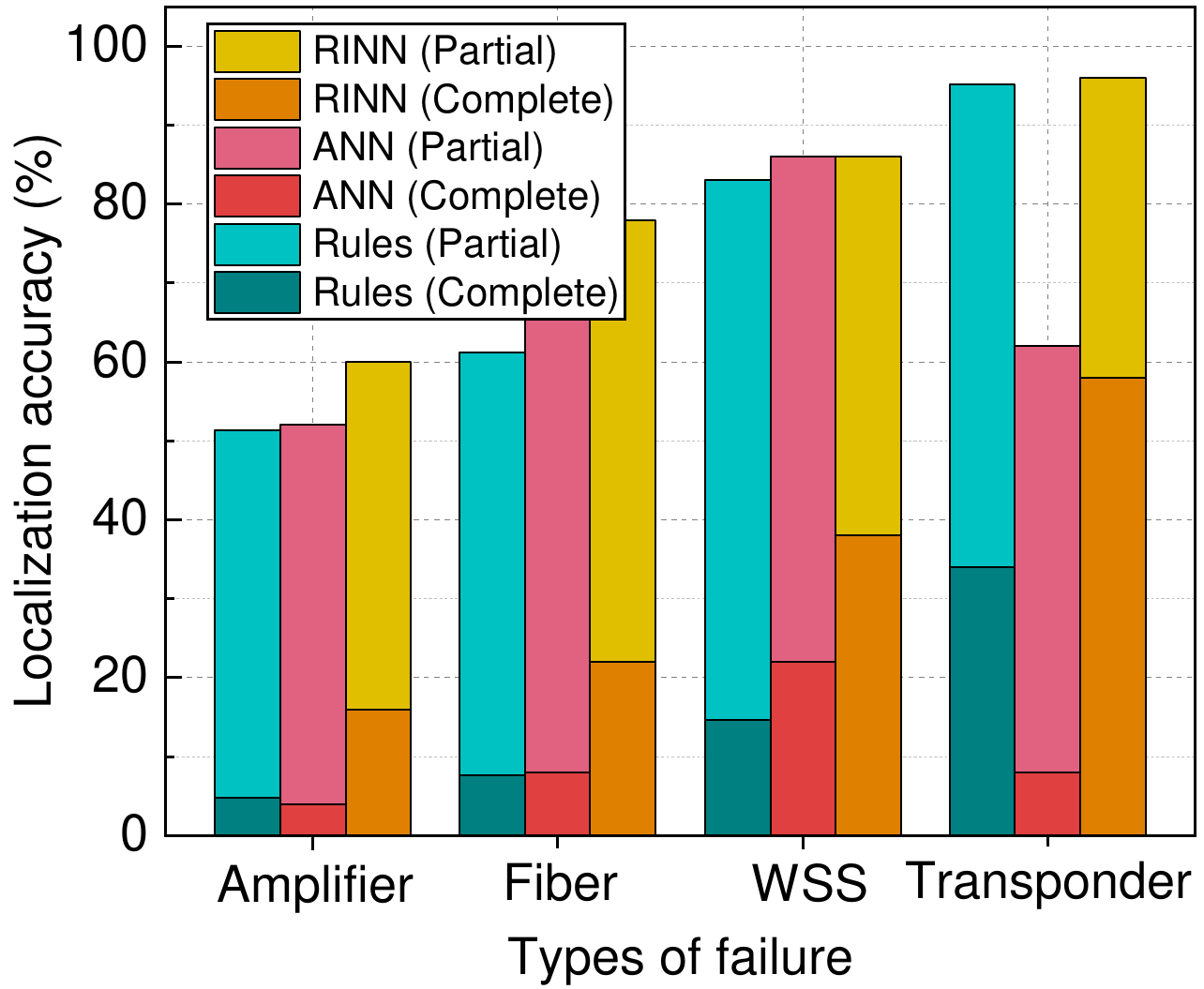}
	\caption{\centering \small 2-failure}
\end{subfigure}
\begin{subfigure}[b]{0.24\textwidth}
	\centering
	\includegraphics[width=\textwidth]{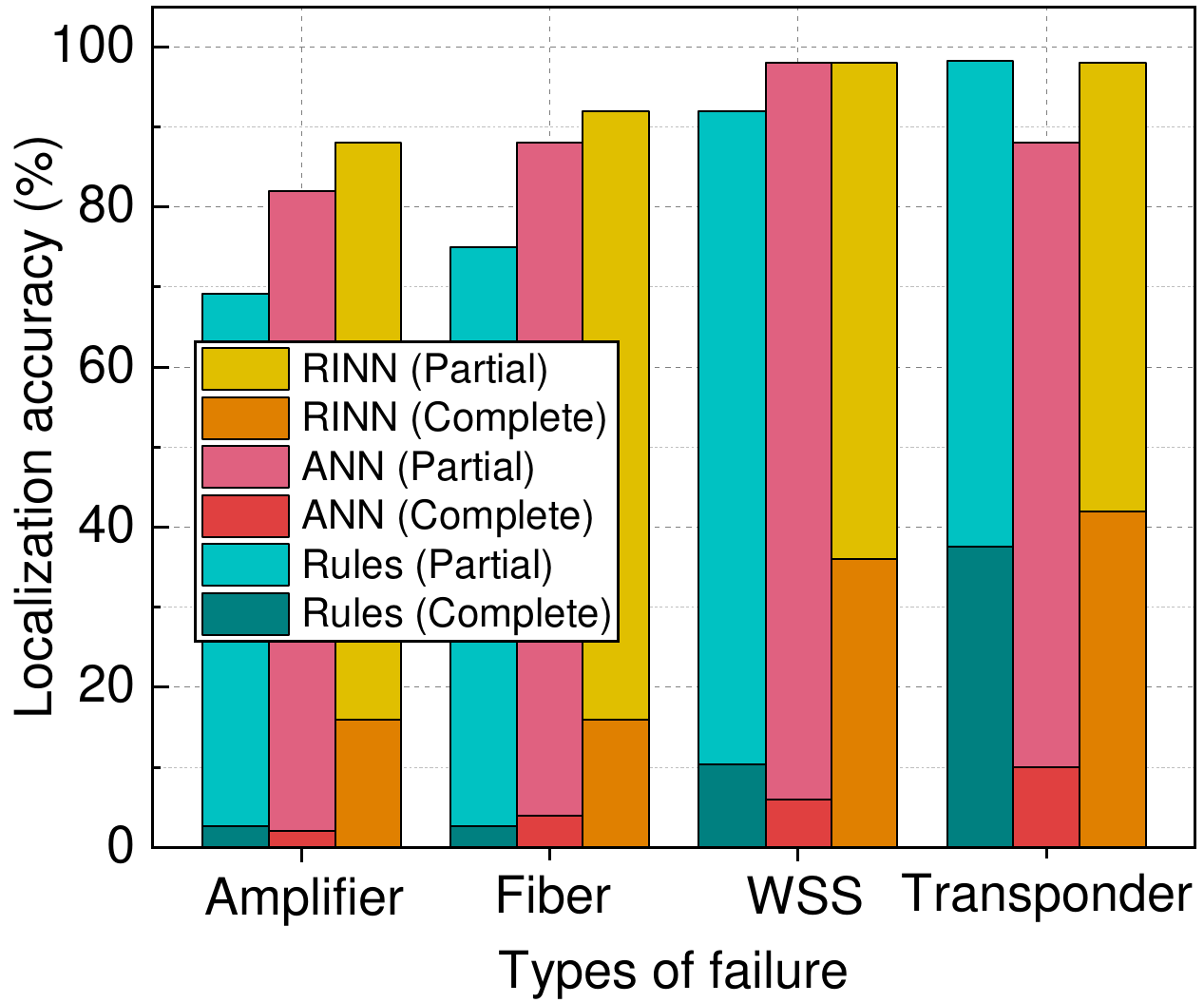}
	\caption{\centering \small 3-failure}
\end{subfigure}
\begin{subfigure}[b]{0.24\textwidth}
	\centering
	\includegraphics[width=\textwidth]{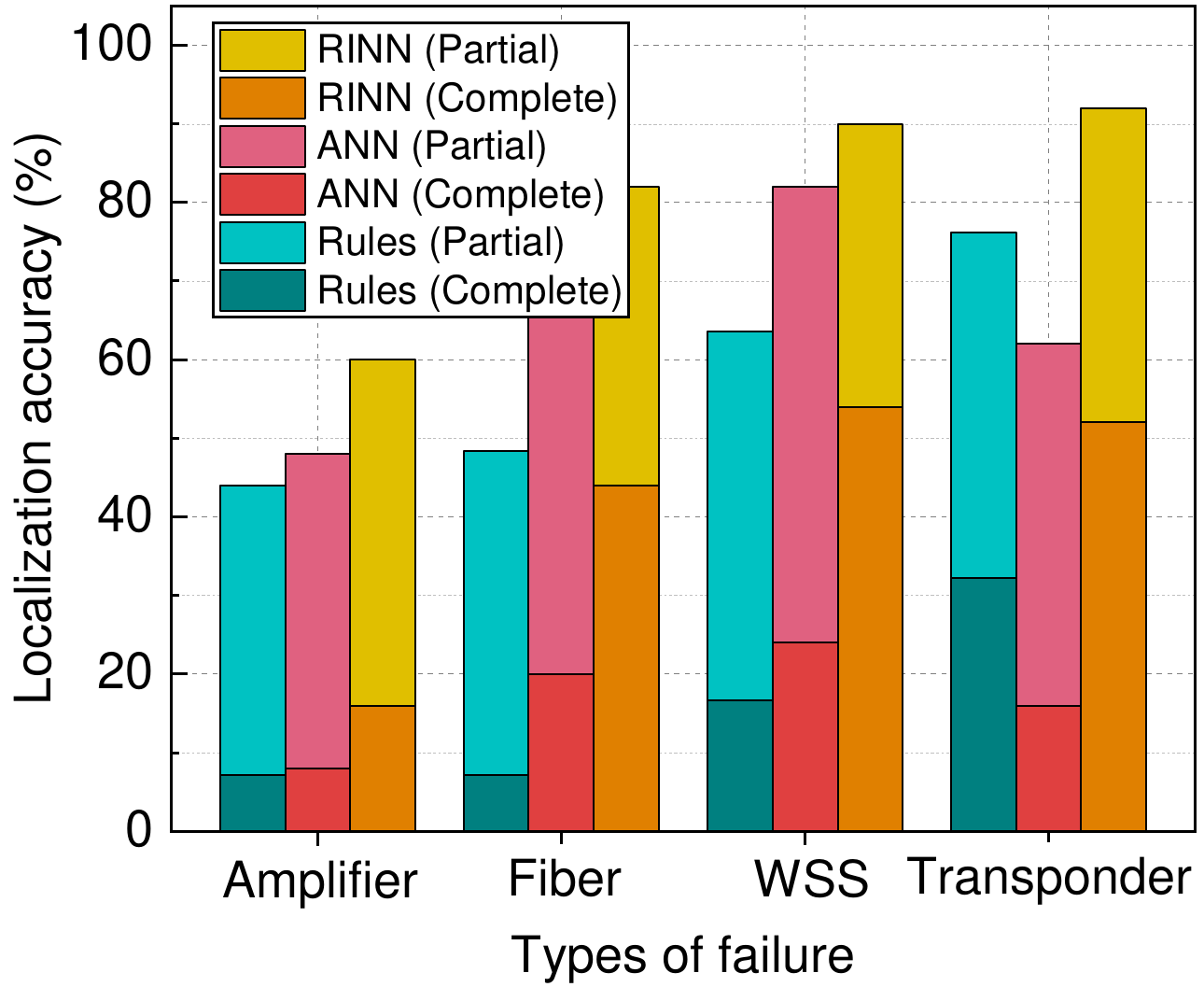}
	\caption{\centering \small mixed-failure}
\end{subfigure}
\caption{Localization accuracy under different types of failure.}
\label{fig:result_type_of_failure}
\end{figure*}

\subsection{Simulation Results}

\subsubsection{Evaluations of rules-based reasoning}
We first evaluate the performance of rules-based reasoning in terms of the ratio of suspected faulty components under different percentages of OPMs. 
As shown in Fig.~\ref{fig:evaluation_rules_based_reasoning} (a), the ratio of suspected faulty components decreases when the percentage of OPMs increases, as the status of more components can be determined with more OPMs. 
For instance, for \textit{1-Failure}, the ratio of suspected faulty components decreases from 26\% to 0\% when the percentage of OPMs increases from 10\% to 100\%. 
Moreover, the ratio of suspected faulty components increases with the number of failures because more LPs are affected by the failure, making it harder to determine the status of components. For instance, the ratio of suspected faulty components for \textit{3-Failure} is up to 1\% higher than that for \textit{1-Failure}. 
Note that the ratio of suspected faulty components for \textit{Mixed-Failure} is close to that for \textit{2-Failure} as the results of \textit{Mixed-Failure} are averaged from failure cases with an average number of failed components equal to 2 (the number of failed components are randomly selected from 1, 2, and 3). 

As shown in Fig.~\ref{fig:evaluation_rules_based_reasoning} (b), we further compare the ratio of suspected faulty components under different number of LPs, where the percentage of OPMs is 60\%.
The ratio of suspected faulty components reduces when the number of LPs increases, as more components are traversed by at least one LP. For instance, the ratio of suspected faulty components reduces from 34\% to 7\% when the number of LPs increases from 40 to 100. 
Moreover, it is more difficult to identify the status of components with a larger number of failed components. 
For instance, the ratio of suspected faulty components increases up to around 1\% in \textit{3-Failure} compared to \textit{1-Failure}. 
The above results indicate that the proposed rules-based reasoning obtains better performance with the increase of the percentage of OPMs and the number of LPs.

\subsubsection{Evaluations of localization accuracy under different percentages of OPMs}
\label{Sec.V-D-2}
We now evaluate the localization accuracy across different percentages of OPMs, 
by generating a distinct training dataset for each percentage of OPM and training an ANN model separately.
In each training dataset, the number of failures is randomly selected from one, two, and three (i.e., mixed-failure), with the failure types selected randomly from all types, and the number of LPs fixed at 100.
We further explain the test dataset used in test procedure.
Each subfigure of Fig.~\ref{fig:result_percentage_of_opm} contains five percentage of OPMs, i.e., 20\% to 100\% (step by 20\%), and there are four subfigures, i.e., \textit{1-failure}, \textit{2-failure}, \textit{3-failure} and \textit{mixed-failure}. 
Therefore, the total number of test dataset is $5 \times 4 = 20$, where each of them contains different percentages of OPMs and/or different number of failures, while the type of failures is randomly selected from all types, and the number of LPs is 100.

Fig.~\ref{fig:result_percentage_of_opm} (a) shows the localization accuracy with only one failure. 
The partial localization accuracy is 0\% for all approaches, as only one component fails, and the failed component is either identified or not. 
The \textit{RINN} always has higher complete localization accuracy than both \textit{ANN} and \textit{Rules} across different percentages of OPMs when the percentage of OPMs is lower than 100\%. 
Specifically, \textit{RINN} outperforms \textit{Rules} and \textit{ANN} by up to 44\% and 30\%, respectively. 
When the percentage of OPMs is 100\%, all methods achieve 100\% localization accuracy, as all the components can be monitored. 
Moreover, when the percentage of OPMs equals 20\%, the complete localization accuracy of \textit{Rules} is around 0.6\% higher than that of \textit{ANN}, showing that \textit{ANN} does not have a good performance with a small number of OPMs. 
When the percentage of OPMs increases (not reaching 100\%), the performance of \textit{ANN} becomes better than \textit{Rules}. For example, when the percentage of OPMs is 80\%, the complete localization accuracy of \textit{ANN} is 36\% higher  than \textit{Rules}.

When the number of failures simultaneously occurring in the network increases, the overall trend for complete localization accuracy remains the same for different approaches, as shown in Fig.~\ref{fig:result_percentage_of_opm} (b)-(d). 
The difference is that with more failures, the partial localization accuracy increases from 0\% in \textit{1-failure}. 
We take the failure scenario of 3-failure as an example. As shown in Fig.~\ref{fig:result_percentage_of_opm}(c), the partial accuracy of \textit{RINN}, \textit{ANN} and \textit{Rules} increases to 78\%, 94\% and 72\%, respectively, when the percentage of OPM is 60\%.
Moreover, the total localization accuracy also increases with the number of failures, meaning that it is easier to identify at least one failed component with a larger number of failed components. 
For instance, as shown in Fig.~\ref{fig:result_percentage_of_opm} (c), the total localization accuracy reaches 100\% for all approaches when the percentage of OPMs equals 80\%.
Finally, the performance of mixed-failure behaves similarly to the cases with two failures and three failures, as shown in Fig.~\ref{fig:result_percentage_of_opm} (d). Specifically, when the percentage of OPM is 60\%, the proposed \textit{RINN} has up to 20\% and 28\% higher complete localization accuracy compared to \textit{ANN} and \textit{Rules}, respectively. Moreover, \textit{RINN} has up to 14\% and 28\% higher total localization accuracy
compared to \textit{ANN} and \textit{Rules}, respectively.

\subsubsection{Evaluations of localization accuracy under different types of failure}
\label{Sec.V-D-3}
Fig.~\ref{fig:result_type_of_failure} shows the localization accuracy for failures of different types of components, where the percentage of OPM is 60\%. 
In this evaluation, we train an ANN model using a training dataset generated with 60\% OPM. The number of failures is randomly chosen from one, two, or three (i.e., mixed-failure), the failure types are randomly selected from all available types, and the number of LPs is set to 100.
In the test procedure, each subfigure of Fig.~\ref{fig:result_type_of_failure} contains four types of failures, and there are four subfigures.
Therefore, the total number of test dataset is $4 \times 4 = 16$, where
each test dataset contains different types of failures and/or different number of failures, where the number of LPs is 100.

For the scenario with one failure, the complete localization accuracy of \textit{RINN} is higher than \textit{ANN} and \textit{Rules} for the failure of all types of components, as shown in Fig.~\ref{fig:result_type_of_failure} (a). 
When the number of failures increases, the total localization accuracy for all types of failure increases, as it is more likely to identify at least one failure with an increased number of failed components, as shown in Fig.~\ref{fig:result_type_of_failure} (b)-(d). 
Moreover, it is worth noticing that the performance of \textit{Rules} is always better than ANN for the failure of the transponder regardless of the number of failed components. 
For instance, the complete localization accuracy and total localization accuracy of \textit{Rules} are around 16\% and 14\% higher than the of ANN for transponder failure with \textit{3-failure}, as shown in Fig.~\ref{fig:result_type_of_failure} (c). 
This is because each transponder is only connected to one LP, and hence only one feature is available for ANN per failure scenario, leading to insufficient features to identify the failure location for transponders. Instead, multiple features are provided to other components such as amplifiers, as these components are likely to be traversed by several LPs, providing more features for ANN. 

\begin{figure}[htb]
\centering
\includegraphics[width=1.0\linewidth]{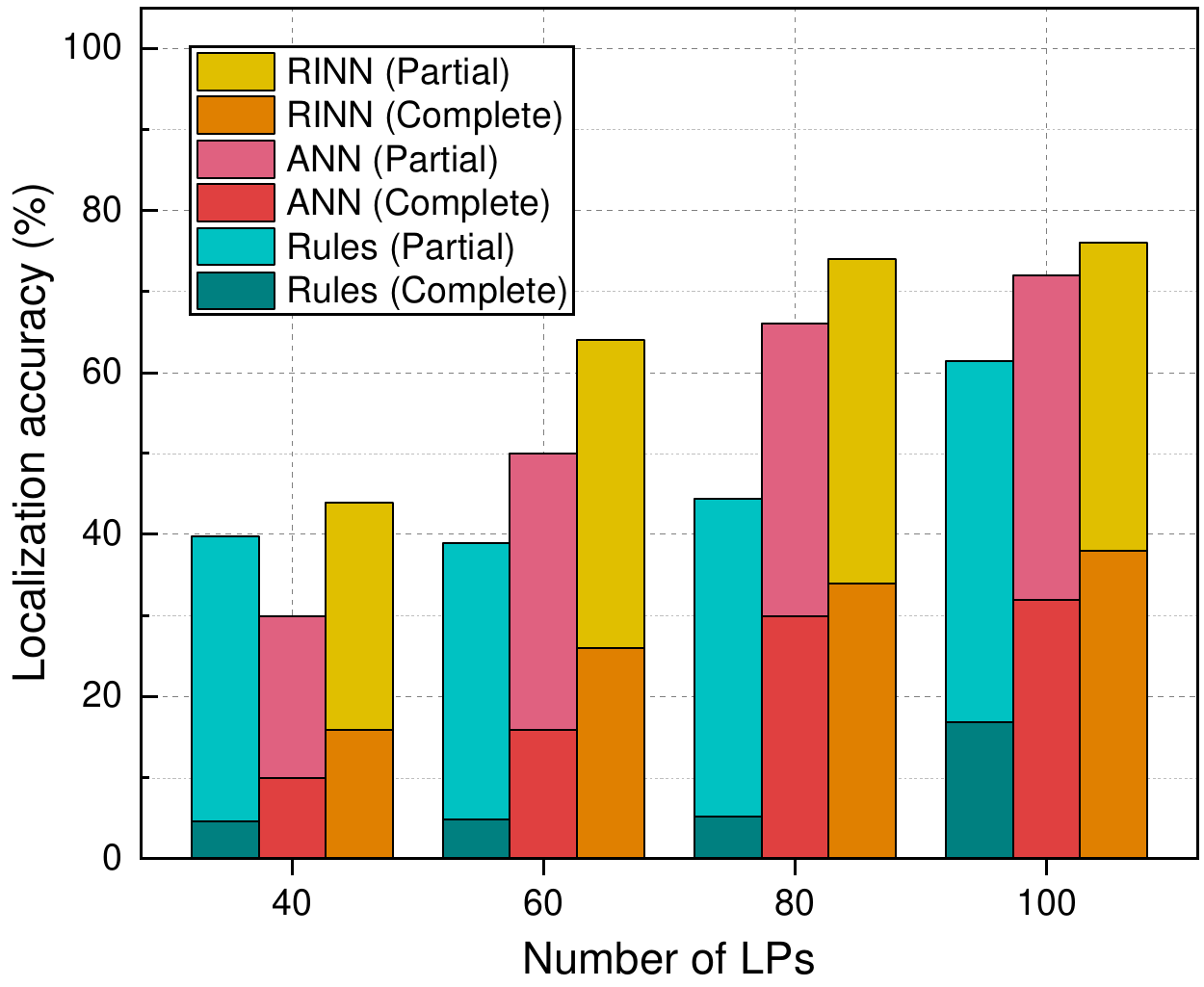}
\caption{Localization accuracy under different number of LPs.}
\label{Fig-result-accuracy-number-of-lp}
\end{figure}

\subsubsection{Evaluation of localization accuracy under different number of LPs}
We now evaluate the localization accuracy under different numbers of LPs. 
This evaluation uses the same ANN model described in Sec.~\ref{Sec.V-D-3}.
As the performance under different number of failures has already been evaluated in Sec.~\ref{Sec.V-D-2} and Sec.~\ref{Sec.V-D-3}, this part only shows the most complicated scenario with the \textit{mixed-failure},
where the percentage of OPMs is 60\%.
Since the number of LPs ranges from 20 to 100 (step by 20) in our evaluation, this work uses 5 test dataset to evaluation all approaches, where each test dataset includes all types of failures.

As shown in Fig.~\ref{Fig-result-accuracy-number-of-lp}, the \textit{RINN} always has higher \textit{complete localization accuracy} and \textit{total localization accuracy} than both \textit{Rules} and \textit{ANN} across different numbers of LPs. 
Specifically, \textit{RINN} outperforms \textit{Rules} and \textit{ANN} by up to around 29\% and 10\%, respectively, in terms of \textit{complete localization accuracy}. 
In addition, \textit{RINN} has up to around 30\% and 14\% higher \textit{total localization accuracy} than \textit{Rules} and \textit{ANN}, respectively. 
Moreover, when the number of LPs equals 40, the \textit{total localization accuracy} of \textit{Rules} is around 10\% higher than that of \textit{ANN}, showing that \textit{ANN} does not have a good performance with a small number of power values. 
When the number of LPs increases, the performance of \textit{ANN} becomes higher than \textit{Rules} in terms of both \textit{complete localization accuracy} and \textit{total localization accuracy}. 
For instance, when the number of LPs equals 100, the \textit{complete localization accuracy} and \textit{total localization accuracy} of \textit{ANN} is up to around 15\% and 11\% than that of \textit{partial localization accuracy}.

\subsubsection{Evaluation of interference time}
\begin{figure}[tb]
\centering
\includegraphics[width=1.0\linewidth]{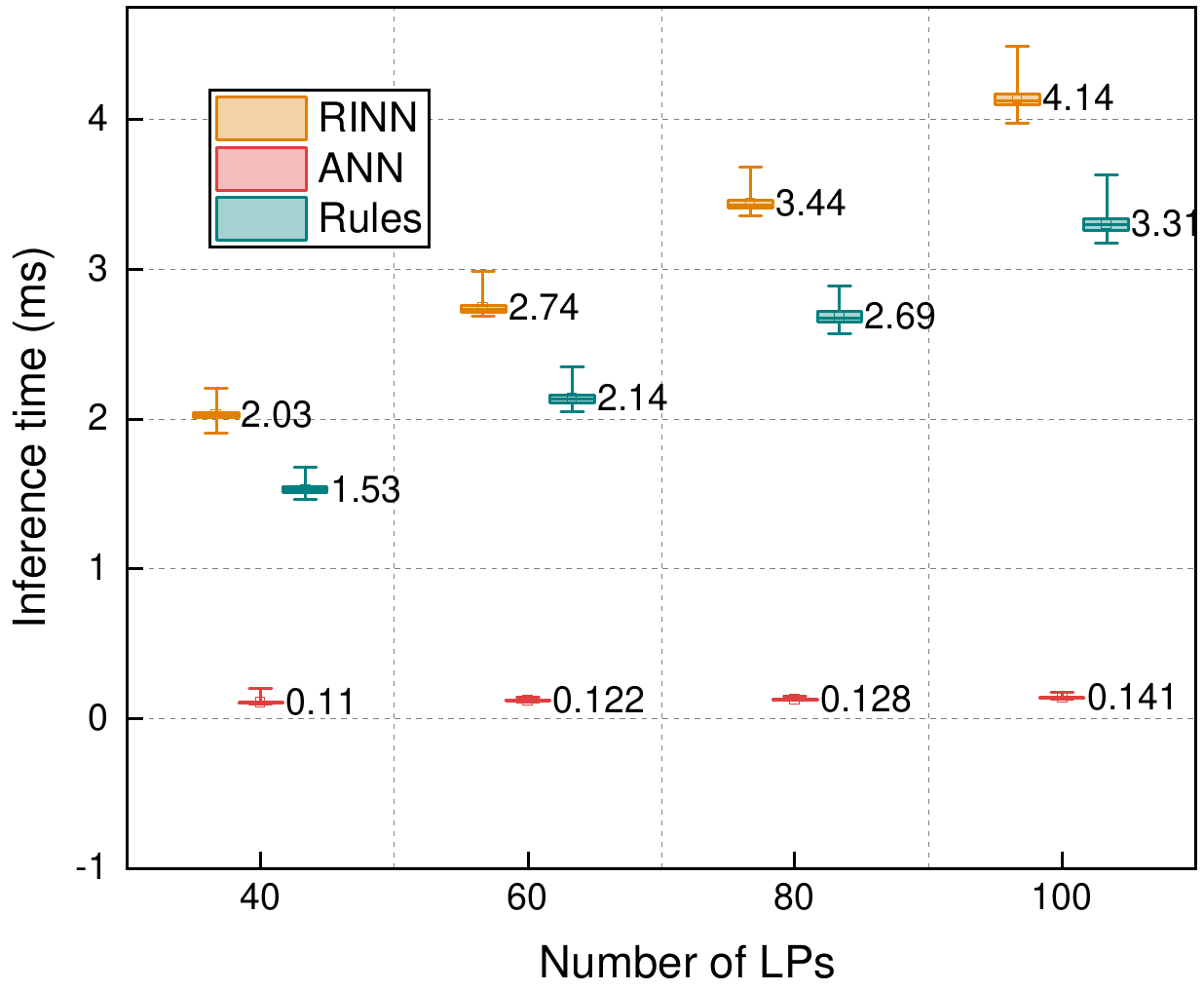}
\caption{Inference time under different number of LPs.}
\label{Fig-result-inference-time}
\end{figure}

We use a same ANN model (described in Sec.~\ref{Sec.V-D-3}) to evaluate the inference time.
As shown in Fig.~\ref{Fig-result-inference-time}, the average inference time for a single failure sample in all approaches increases when the number of LPs increases. Although RINN has the largest inference time, the highest average inference time in our evaluations is still less than 4.14 ms, showing that our proposed solution is swift in identifying faulty components in the network. 

\subsubsection{Evaluation of RINN convergence}

\begin{figure}[tb]
\centering
\includegraphics[width=1.0\linewidth]{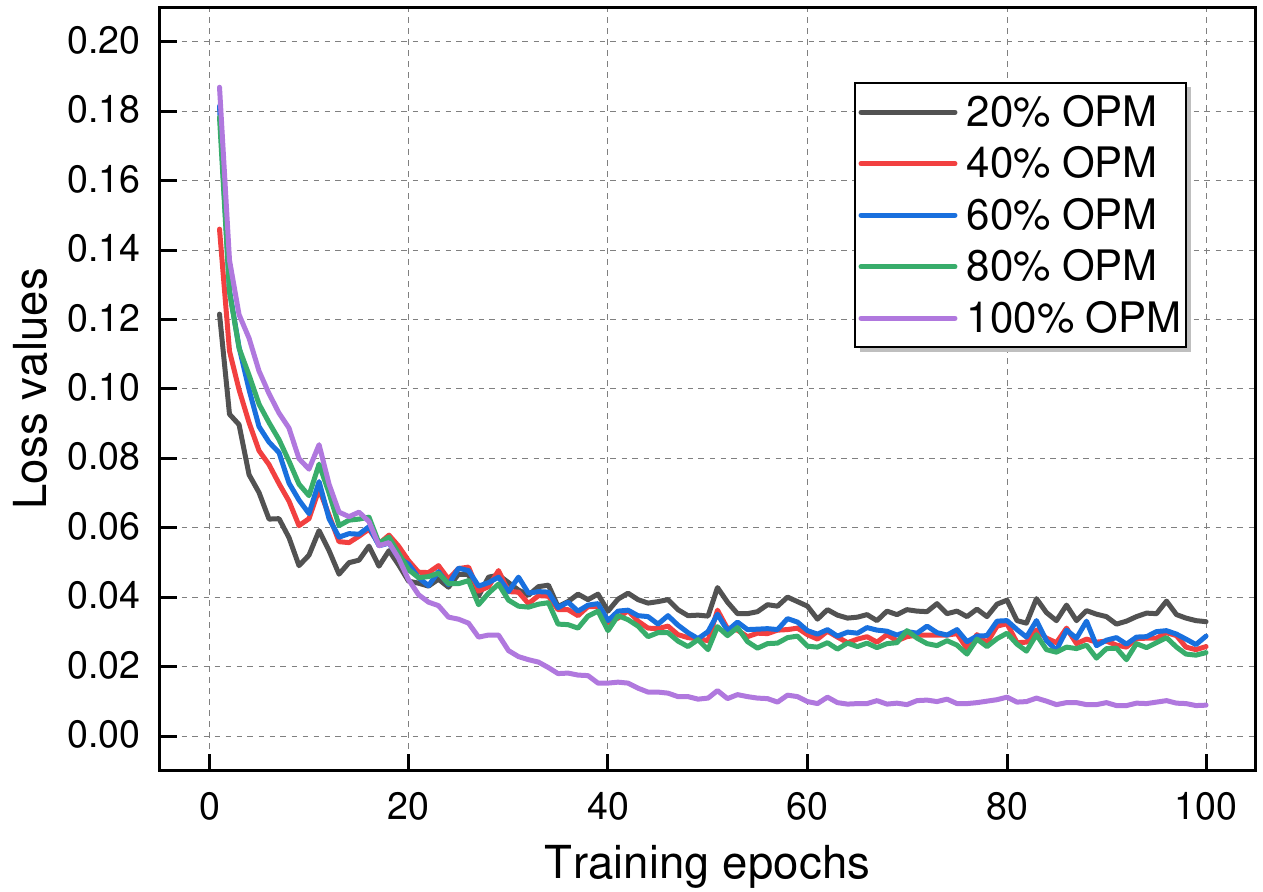}
\caption{Loss values under different training epochs.}
\label{Fig-result-NN-convergence}
\end{figure}

We further evaluate the convergence procedure of the proposed RINN using a same ANN model described in Sec.~\ref{Sec.V-D-3}. Fig. \ref{Fig-result-NN-convergence} shows the cross-entry loss values\footnote{As mentioned in Sec. \ref{sec:subsec_neural_network}, we perform binary classification for each suspected faulty component, and the output status is failure or non-failure. The loss values are calculated according to Eq. (\ref{equ-cross-entry}).} under different number of training epochs, where the number of LPs is 100. The results indicate that the training loss values decrease rapidly with the training epochs, and then converge to stable values. For example, the loss values converge to 0.008 when the percentage of OPMs is 100\%.
Furthermore, the loss values decrease when the percentage of OPMs increases, since the more monitoring features can be obtained with more OPMs.

\subsection{Experimental Demonstrations and Results}
\begin{figure}[tb]
\centering
\includegraphics[width=1.0\linewidth]{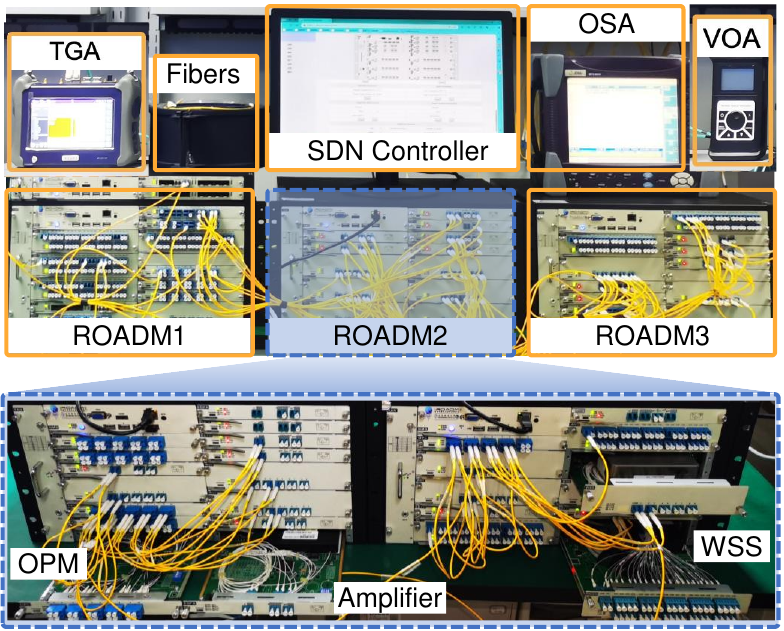}
\caption{Experimental testbed with a 3-node ROADM-based optical network.}
\label{Fig-testbed}
\end{figure}

\begin{figure}[tb]
\centering
\includegraphics[width=1.0\linewidth]{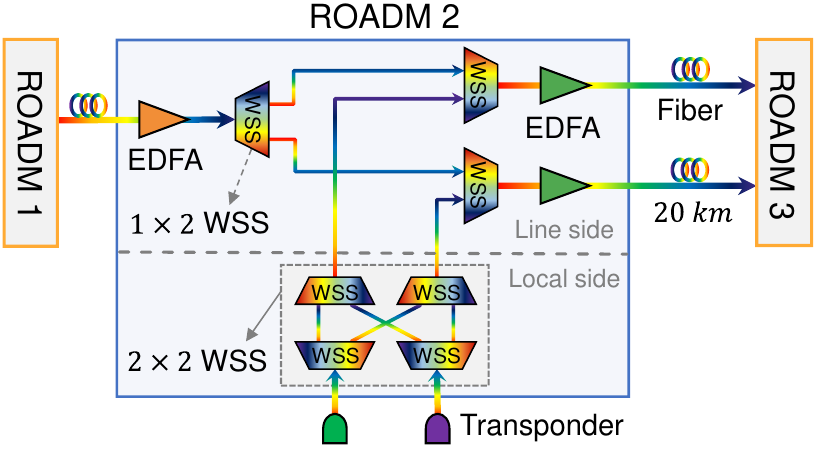}
\caption{Experimental scenario with a 3-node ROADM-based optical network.}
\label{Fig-testbed-scenario}
\end{figure}

We perform experimental demonstrations on a real 3-node ROADM-based testbed, as shown in Fig.~\ref{Fig-testbed}. 
This testbed is composed of three ROADM nodes interconnected by fibers. 
Each ROADM node consists of a complex internal composition, i.e., amplifiers, WSS, and transponders, as shown in bottom of Fig.~\ref{Fig-testbed}. 
This testbed is controlled by an SDN controller, which can perform wavelength allocation for WSSs, and control amplifier gains for amplifiers.
Moreover, the SDN controller can automatically gather the power values from OPMs. 
The live traffic of the optical network can be generated via \textit{traffic generator and analyzer} (TGA). 
The failure can be inserted through \textit{variable optical attenuator} (VOA), which is able to insert loss in any location of an LP. The status of each LP can be monitored using \textit{optical spectrum analyzer} (OSA). 
Fig.~\ref{Fig-testbed-scenario} shows the corresponding experimental scenario. On the local side, the $2\times2$ WSS is composed of four $1\times2$ WSS\footnote{Due to the lack of experimental components, we do not have $m\times n$ WSS. Instead, we use multiple $1\times k$ WSSs to form an $m\times n$ WSS.}, and several transponders are connected with WSS to inject live traffic.
On the line side, several EDFA and WSS are used to gain amplification and wavelength switching, respectively.

\begin{figure*}[htb]
\centering
\includegraphics[width=1.0\linewidth]{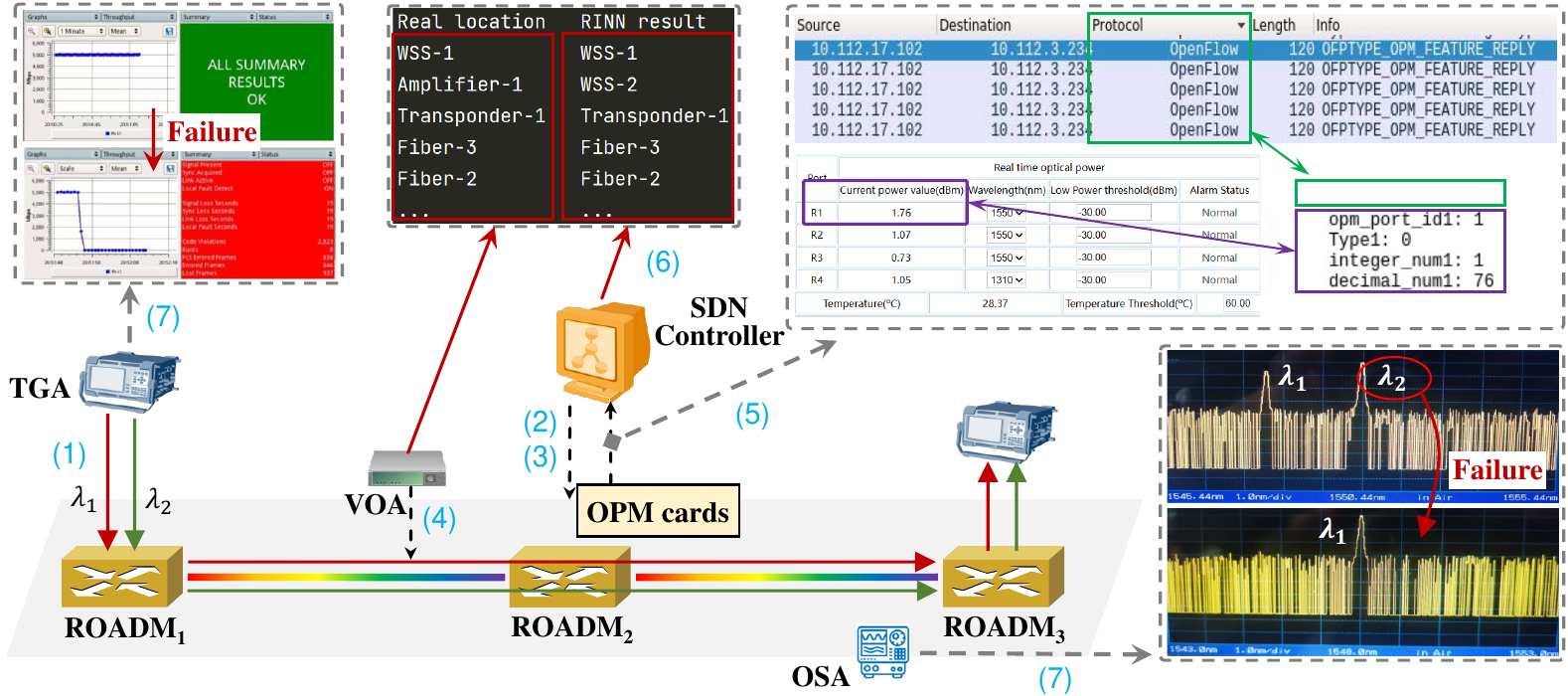}
\caption{Experimental demonstration procedures and results.}
\label{Fig-experimental-results}
\end{figure*}

\begin{figure}[tb]
\centering
\includegraphics[width=1.0\linewidth]{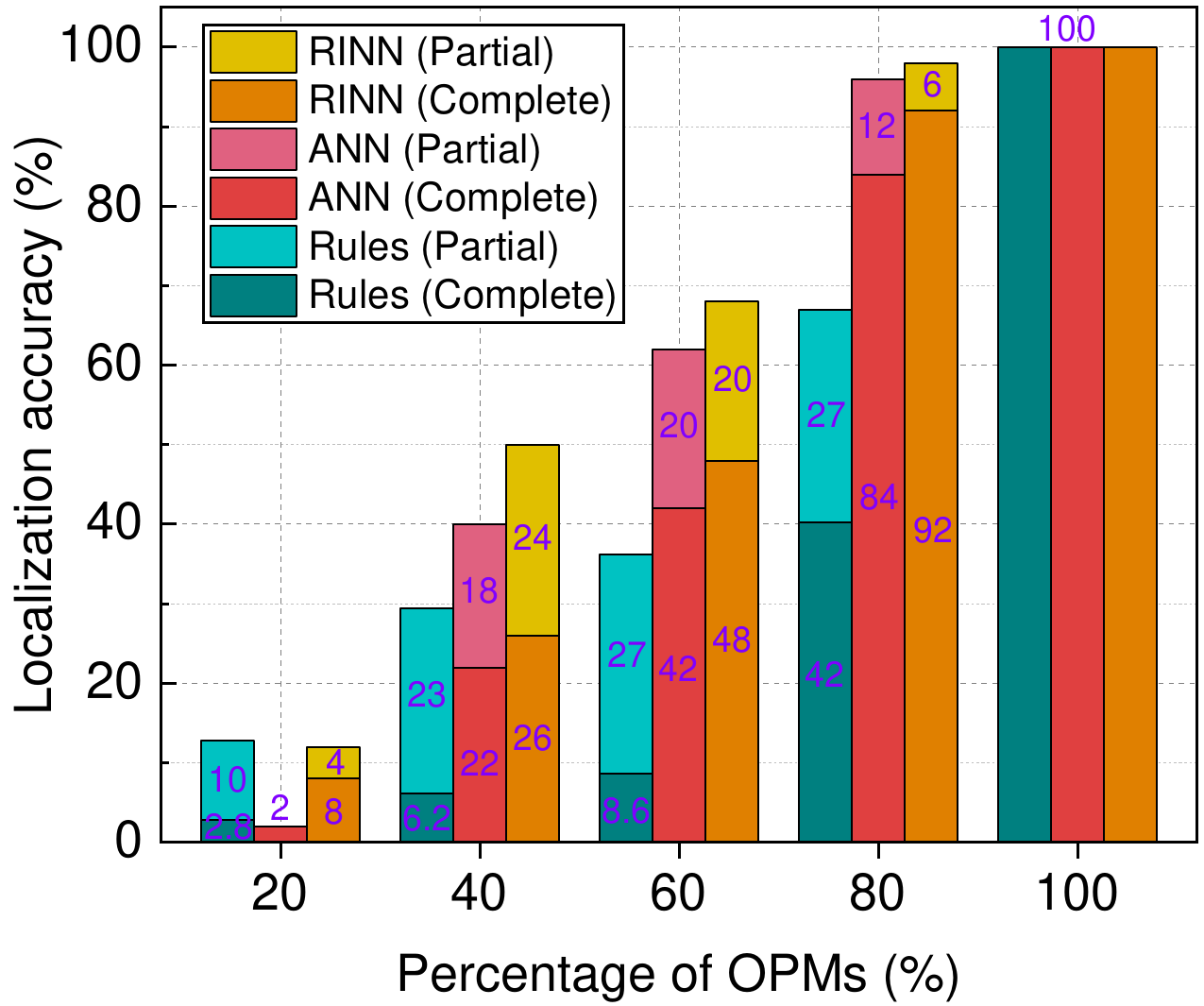}
\caption{Localization accuracy vs. different percentage of OPMs in a testbed.}
\label{Fig-result-experimental-values}
\end{figure}

The experimental demonstration procedures are illustrated in Fig.~\ref{Fig-experimental-results}. 
The procedures in the figure are ordered with blue numbers. 
The first step is to load traffic through TGA. Then the SDN controller performs RFWA (step 2) and configures WSSs and amplifiers to set connections for all the LPs (step 3). 
After setting the connections, VOA is deployed to generate failures, such as WSS-1 and Fiber-3 shown in top middle of Fig. \ref{Fig-experimental-results} (step 4). 
Specifically, the components directly connected to the VOA are considered to be failed components. 
After generating failures, the power values are reported from OPMs to the SDN controller (step 5).
For example, the power value of one OPM (equal to $1.76$ $dBm$ and marked with a purple box) is shown in the table on the top right of Fig.~\ref{Fig-experimental-results}. 
The RINN framework determines the failed components based on the monitored OPM values (step 6), and the results are shown in top middle of Fig. \ref{Fig-experimental-results}.
Finally, we can monitor the live traffic and LP status using TGA and OSA, respectively (step 7).
In this example, the live traffic is changing from $5$ $Gbps$ to $0$ $Gbps$, and one LP request is rejected (i.e., $\lambda_2$), as shown in top left and bottom right of Fig. \ref{Fig-experimental-results}, respectively.

We collect 200 training samples and 100 test samples, respectively. These samples are used to train and test all algorithms, respectively.
As shown in Fig.~\ref{Fig-result-experimental-values}, we also present the localization accuracy under different percentage of OPMs, where the number of failures are randomly selected from "1" or "2". Unlike Fig.~\ref{fig:result_percentage_of_opm}, the experimental data in Fig.~\ref{Fig-result-experimental-values} are collected from a real testbed.
The experimental results indicate that the proposed RINN achieves higher localization accuracy than baseline methods before the percentage of OPMs reaches 100\%.

Based on the above simulation and experimental evaluations, we summarize the reasons why the proposed RINN outperforms the \textit{Rules} benchmark. Firstly, there is a complex dependency relationship between continuous power values and multiple failure locations. The \textit{Rules} benchmark only utilizes the pre-defined rules, and struggles to effectively uncover the dependency relationship.
Secondly, the power values show an oscillatory and changeable nature in time. Since the proposed \textit{RINN} is a data-driven method, it is more suitable for simultaneously analyzing the varying power values.
Finally, as shown in Fig.~\ref{Fig-pipeline-RINN}, the proposed \textit{RINN} incorporates the benefits of both rules-based reasoning and ANN models.
Due to the these reasons, the proposed \textit{RINN} achieves higher localization accuracy than the \textit{Rules} benchmark.

\section{Conclusion}
\label{section-conlusion}
This paper investigated the problem of multi-failure localization of inter-/intra-node components in high-degree ROADM-based optical networks. 
We proposed a rules-informed neural network (RINN) for multi-failure localization, which combines the strengths of rules-based reasoning for a small number of OPMs and the strengths of artificial neural networks (ANN) for a large number of OPMs. 
Our extensive simulations and experimental demonstrations validate the superiority of the RINN algorithm, showcasing an enhancement in localization accuracy of up to 20\% and 18\% over \textit{Rules} and \textit{ANN}, respectively, while maintaining a practical inference time of around 4.14 ms. In the future, we plan to further investigate our propose solution when applied to other ROADM architectures like Clos-based high-degree ROADM and cascaded high-degree ROADM. \textcolor{black}{
The proposed RINN solution requires a significant number of training samples, which presents a critical challenge in practical scenarios. Some methods can address the issue of sample scarcity, e.g., digital twins and transfer learning. Moreover, the RINN requires to be trained on data belonging a specific network deployment, and generalization of the RINN to multiple topologies remains an open research challenge. We plan to explore these approaches in future works.
}

\bibliographystyle{IEEEtran}
\bibliography{references}


\begin{IEEEbiography}[{\includegraphics[width=1in,height=1.25in,clip,keepaspectratio]{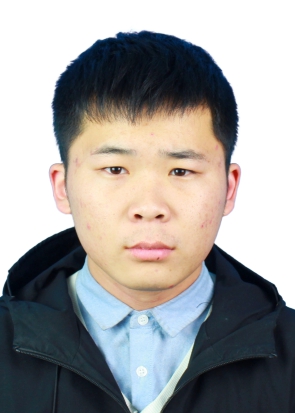}}]{Ruikun Wang}
(Graduate Student Member,~IEEE) is currently a Ph.D student at Beijing University of Posts and Telecommunications, Beijing, China. He was a Joint-Supervised Ph.D. student with the Politecnico di Milano, Milano, Italy. His current research interests are in the fields of Machine-Learning-based failure management and resource optimization.
\end{IEEEbiography}

\begin{IEEEbiography}[{\includegraphics[width=1in,height=1.25in,clip,keepaspectratio]{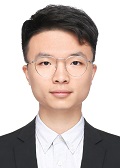}}]{Qiaolun Zhang}
(Graduate Student Member,~IEEE) is currently a Ph.D student at Politecnico di Milano, Milan, Italy. 
He was a optical networking assistant researcher at Nokia Bell Labs France in 2023. 
His current research interests are in the fields of optimization and machine-learning approaches for resource allocation in optical networks. 
\end{IEEEbiography}

\begin{IEEEbiography}[{\includegraphics[width=1in,height=1.25in,clip,keepaspectratio]{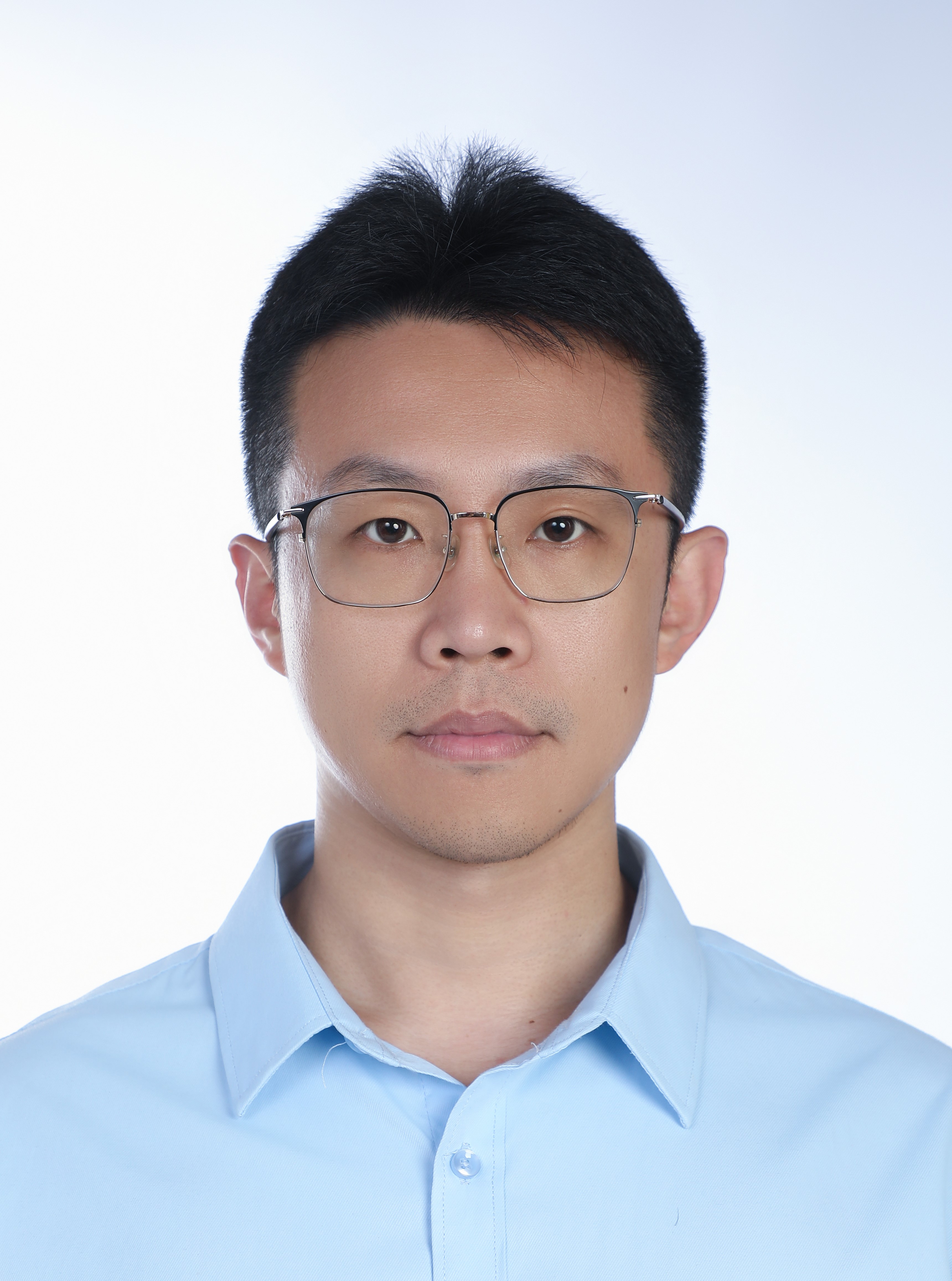}}]{Jiawei Zhang}
(Member, IEEE) received his Ph.D. degree from State Key Lab of Information Photonics and Optical Communication at Beijing University of Posts and Telecommunications (BUPT), Beijing, China, in 2014. He is currently a professor of State Key Lab of Information Photonics and Optical Communication at BUPT. His research interests are optical network architecture, algorithms and system implementation technologies. In the past five years, he has authored/correspond authored more than 30 OFC/ECOC papers in which five of them were top-scored. He served as the technical program chair/member of many conferences, such as ACP 2021/2022, ONDM2022, DRCN2020 et al. He is now serving as a TPC member for the session of optical architectures and software-defined control for metro and more optical networks (N3) in OFC2025.
\end{IEEEbiography}

\begin{IEEEbiography}[{\includegraphics[width=1in,height=1.25in,clip,keepaspectratio]{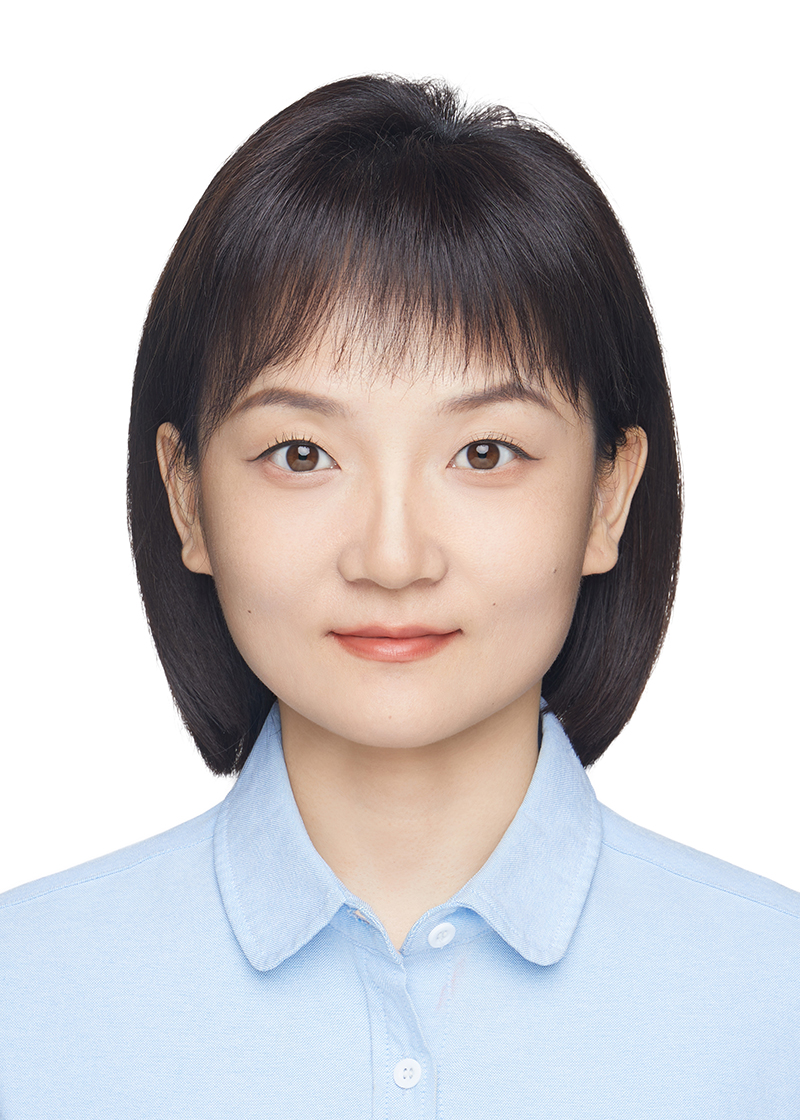}}]{Zhiqun Gu}
(Member,~IEEE) received the Ph.D. degree from the Beijing University of Posts and Telecommunications (BUPT), Beijing, China. She is currently an Associate Professor with BUPT. Her research interests include intelligent optical networks, free space optical communication, and satellite optical networks.
\end{IEEEbiography}

\begin{IEEEbiography}[{\includegraphics[width=1in,height=1.25in,clip,keepaspectratio]{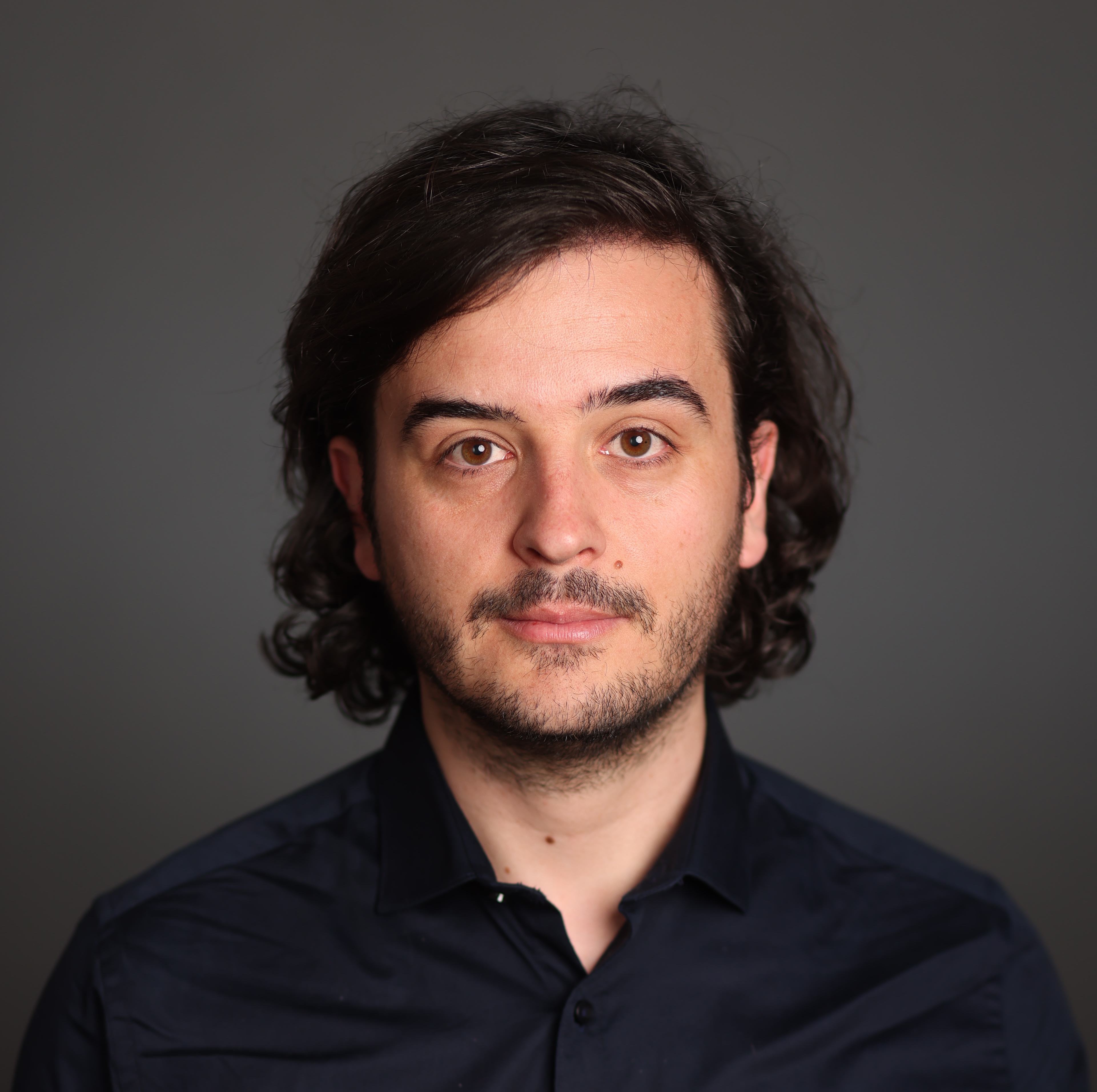}}]{Memedhe Ibrahimi}
(Member, IEEE) is currently an Assistant Professor with the Department of Electronics, Information and Bioengineering at Politecnico di Milano. Dr. Ibrahimi received his Ph.D. degree in Information Engineering from Politecnico di Milano, in 2022. His main research interests include cross-layer design and optimization of long‑haul and metro‑area optical networks, and the application of Machine Learning in the context of communication networks. Dr. Ibrahimi is the author of more than 30 papers published in international journals and conference proceedings, 3 book chapters, and co-winner of a best paper award. 
\end{IEEEbiography}

\begin{IEEEbiography}[{\includegraphics[width=1in,height=1.25in,clip,keepaspectratio]{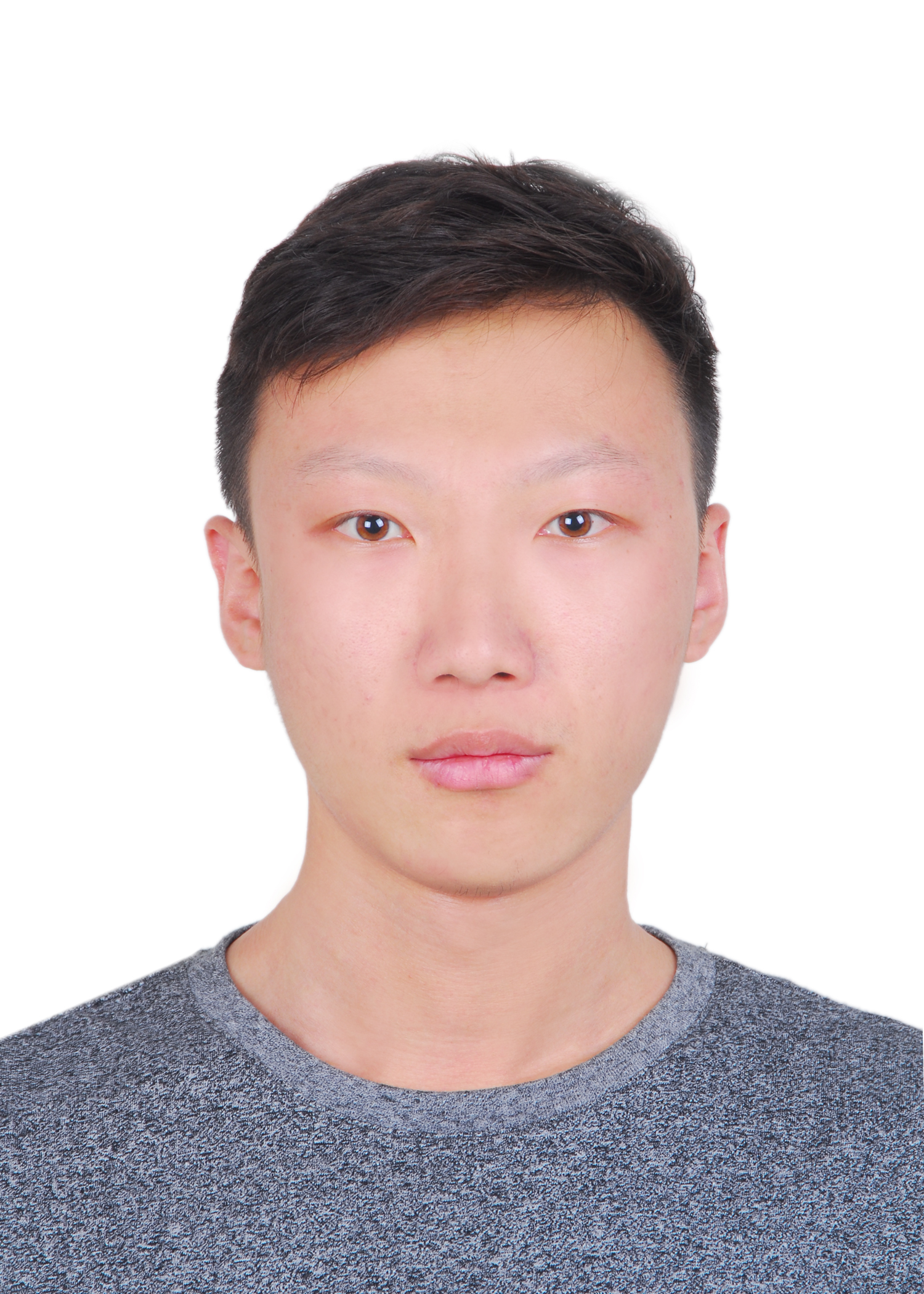}}]{Hao Yu}
(Member, IEEE) received the B.S. and Ph.D. degrees in communication engineering from the Beijing University of Posts and Telecommunications (BUPT), Beijing, China, in 2015 and 2020. He was also a Joint Supervised Ph.D. Student at the Politecnico di Milano, Milano, Italy from 2018 to 2019. He was also a Postdoctoral Researcher in the Department of Information and Communications Engineering, Aalto University, Finland from 2021 to 2022, and in the Center of Wireless Communications, University of Oulu, Finland from 2022 to 2024. Currently, he is working in the ICTFicial Oy, Finland as a senior researcher. His research interests include network automation, SDN/NFV, edge intelligence, time-sensitive networks, and 6G deterministic networking.
\end{IEEEbiography}

\begin{IEEEbiography}[{\includegraphics[width=1in,height=1.25in,clip,keepaspectratio]{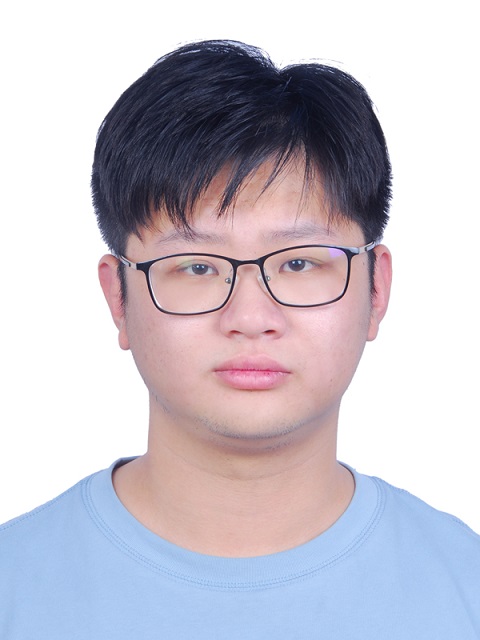}}]{Bojun Zhang}
received the B.S. degree in communication engineering from Beijing University of Posts and Telecommunications in 2020. He is currently a Ph.D student in Beijing University of Posts and Telecommunications, Beijing, China. His current research interests are in the fields of optical network control and resource management, high-degree optical node architecture, etc.
\end{IEEEbiography}

\begin{IEEEbiography}[{\includegraphics[width=1in,height=1.25in,clip,keepaspectratio]{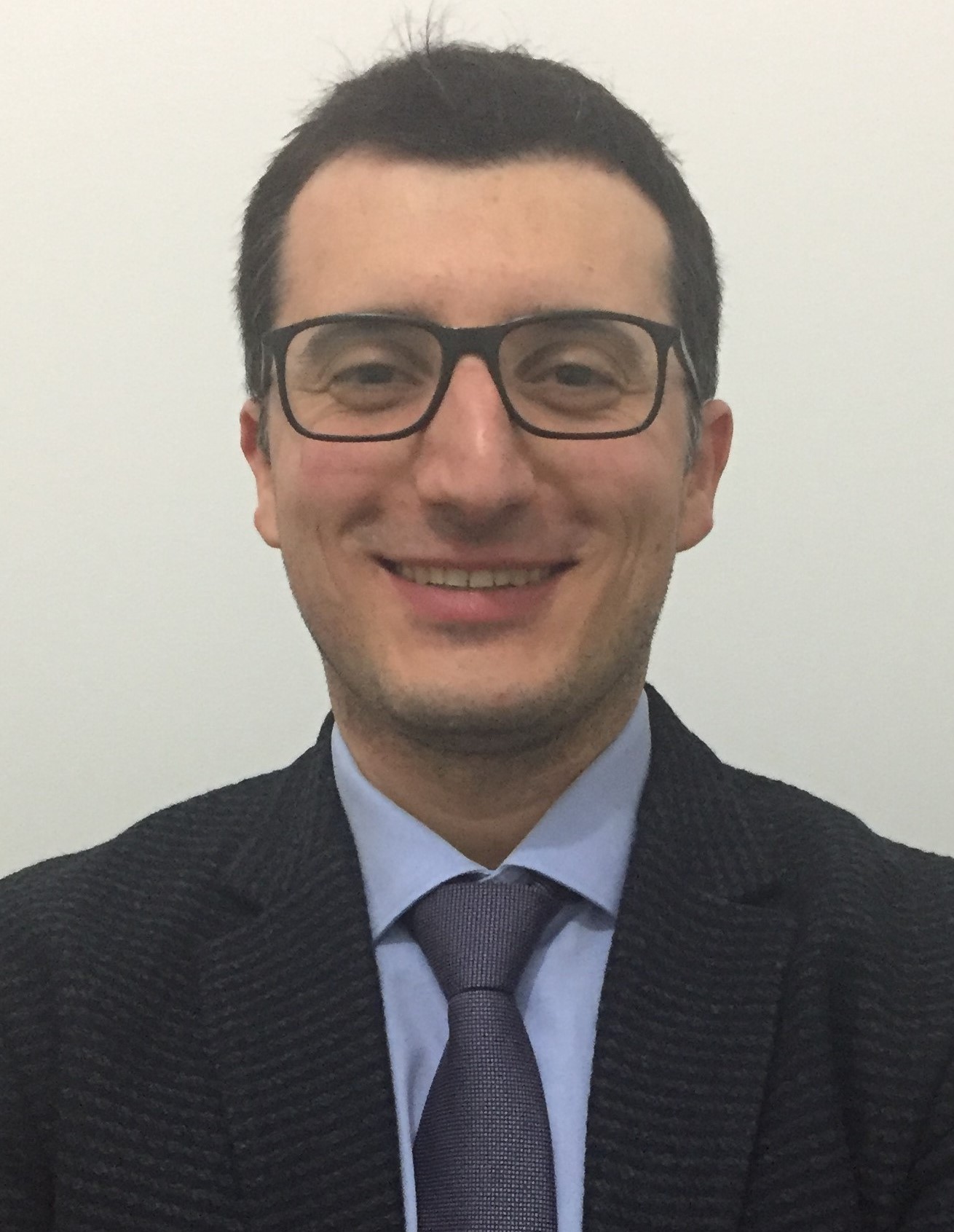}}]{Francesco Musumeci} (S’11-M’12-SM’23) received the Ph.D. degree in Information Engineering from Politecnico di Milano, Italy, in 2013, where he is currently Associate Professor with the Department of Electronics, Information and Bioengineering. His current research interests include Machine-Learning-assisted networking, design and optimization of communication networks, network disaster resilience, and converged space-ground network infrastructures. Dr. Musumeci is author of more than 130 papers published in international journals and conference proceedings, 3 book chapters and 1 patent in the area of communication networks, and is co-winner of three best paper awards from IEEE sponsored conferences.  
\end{IEEEbiography}

\begin{IEEEbiography}[{\includegraphics[width=1in,height=1.25in,clip,keepaspectratio]{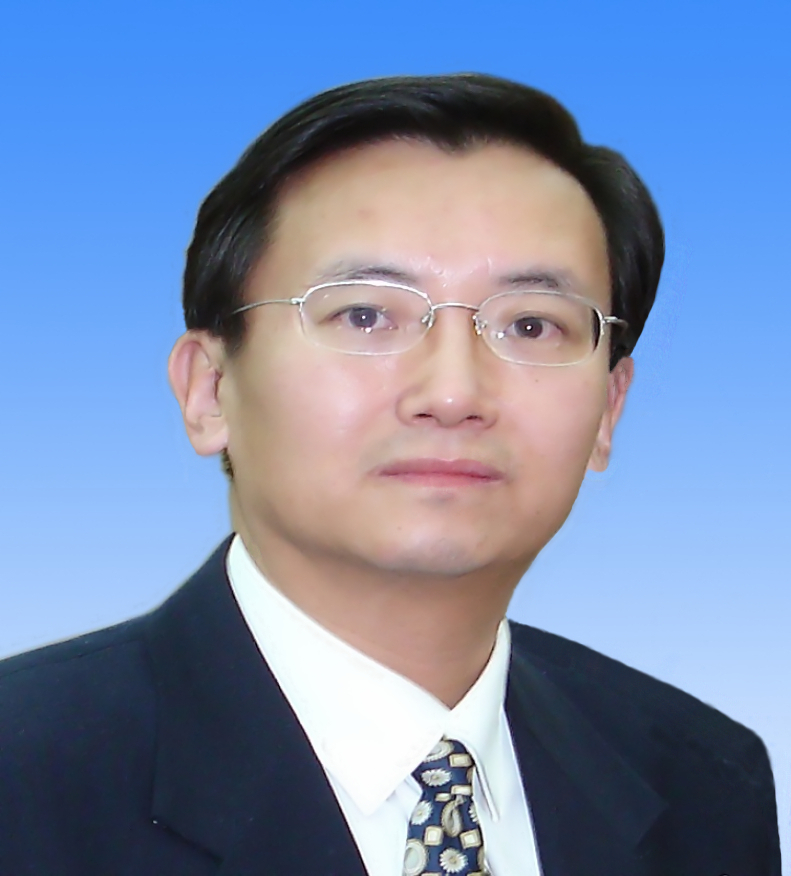}}]{Yuefeng Ji}
(Senior Member, IEEE) received the Ph.D. degree from the Beijing University of Posts and Telecommunications (BUPT), Beijing, China. He is currently a Professor with the State Key Laboratory of Information Photonics and Optical Communications, School of Information and Communication  Engineering, BUPT. His research interests include broadband communication
networks and optical communications.
\end{IEEEbiography}

\begin{IEEEbiography}[{\includegraphics[width=1in,height=1.25in,clip,keepaspectratio]{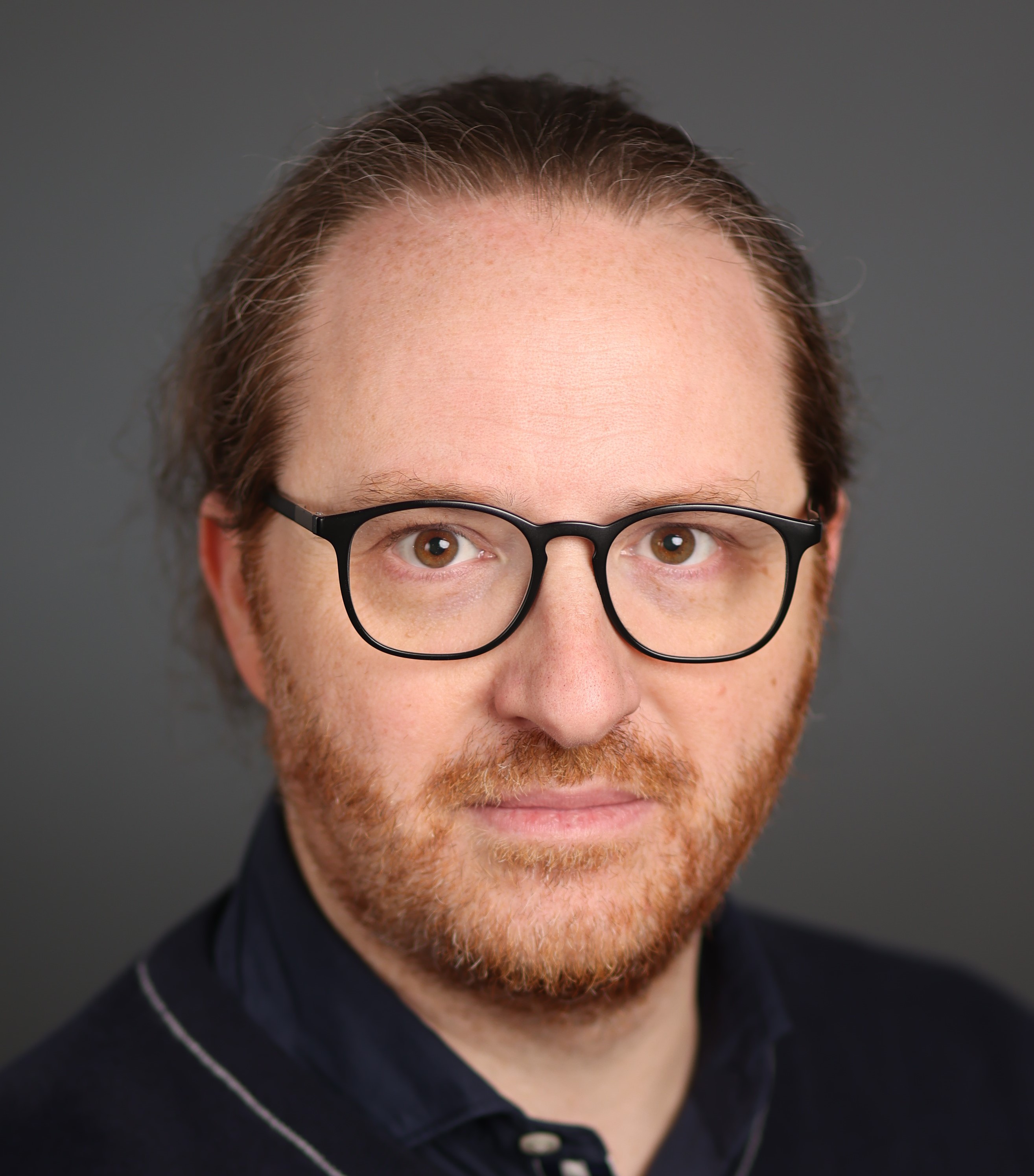}}]{Massimo Tornatore}
(Fellow, IEEE) is currently a Full Professor in the Department of Electronics, Information, and Bioengineering, Politecnico di Milano. He has also held appointments as Adjunct Professor at University of California, Davis, USA and 
as Visiting Professor at University of Waterloo, Canada. 
His research interests include performance evaluation, optimization and design of communication networks (with an emphasis on the application of optical networking technologies), network virtualization, network reliability, and machine learning application for network management. In these areas, he co-authored more than 500 peer-reviewed conference and journal papers (with 21 best paper awards), 2 books, and 3 patents.
He is a member of the Editorial Board, among others, of IEEE Communication Surveys and Tutorials, IEEE Transactions on Networking, IEEE Transactions on Network and Service Management. 
\end{IEEEbiography}

\end{document}